\documentclass[conference]{IEEEtran}

\usepackage{caption}
\usepackage{subcaption}

\usepackage{comment}
\usepackage{amssymb}
\usepackage{amsmath}
\usepackage{amsfonts}
\usepackage{mathrsfs}
\usepackage{engord}
\usepackage[dvips]{graphicx}
\usepackage{bbm}
\usepackage{threeparttable}
\usepackage{booktabs}
\usepackage{epsfig}
\usepackage{color}
\usepackage{graphicx}
\usepackage{multirow}
\usepackage{cite}
\usepackage{epstopdf}

\usepackage{framed}
\usepackage{url}

\usepackage{color}
\usepackage{algorithm2e}

\makeatletter
\newcommand{\removelatexerror}{\let\@latex@error\@gobble}
\makeatother

\newtheorem{thm}{Theorem}

\newtheorem{lem}{Lemma}

\newtheorem{ex}{Example}
\newtheorem{cor}{Corollary}

\newtheorem{rmk}{Remark}

\begin{document}

%
\title{Layered Synthesis of Latent Gaussian Trees}

\author{\IEEEauthorblockN{Ali Moharrer,
Shuangqing Wei,
George T. Amariucai, 
and Jing Deng}}
\maketitle
\footnotetext[1]{A. Moharrer, and S. Wei are with the school of Electrical Engineering and Computer Science, Louisiana State University, Baton Rouge, LA 70803, USA (Email: amohar2@lsu.edu, swei@lsu.edu). 

G. T. Amariucai is with the department of Electrical and Computer Engineering, Iowa State University, Ames, IA, USA (Email: gamari@iastate.edu). 

J. Deng is with the department of Computer Science, University of North Carolina at Greensboro, Greensboro, NC, USA (Email: jing.deng@uncg.edu).

Part of the results was presented in the 54th Annual Allerton Conference on Communication, Control, and Computing, Sept. 2016.

This material is based upon work supported in part by the National Science
Foundation under Grant No. 1320351.}



\begin{abstract}
A new synthesis scheme is proposed to generate a random
vector with prescribed joint density that induces a (latent) Gaussian
tree structure. The quality of synthesis is shown by vanishing
total variation distance between the synthesized and desired statistics.
The proposed layered and successive synthesis scheme relies on the learned
structure of tree to use sufficient number of common random variables
to synthesize the desired density. We characterize the achievable rate
region for the rate tuples of multi-layer latent Gaussian tree, through
which the number of bits needed to synthesize such Gaussian joint density
are determined.  The random sources used in our algorithm are the latent
variables at the top layer of tree, the additive independent Gaussian
noises, and the Bernoulli sign inputs that capture the ambiguity of correlation
signs between the variables. We have shown that such ambiguity can further help in reducing the synthesis rates for the underlying Gaussian trees.

\end{abstract}

\section{Introduction}

Consider the problem of simulating a random vector with prescribed
joint density.  
Such \textit{generative modeling} can be implemented by generating an appropriate number
of random input bits (by relying on a random source) to a stochastic channel whose output vector has its empirical statistics meeting the desired one measured by a given metric.
Generative models have many applications ranging from probabilistic programs \cite{kulkarni} to economics \cite{gourieroux}, physics \cite{cameron} and computer vision \cite{jin2006}.

We aim to address such synthesis problem for a case where the prescribed
output statistics induces a (latent) \textit{Gaussian tree} structure,
i.e., the underlying structure is a tree and the joint density of the
variables is captured by a Gaussian density.  The Gaussian graphical
models are widely studied in the literature, because of a direct
correspondence between conditional independence relations occurring
in the model with zeros in the inverse of covariance matrix, known as
the \textit{concentration matrix}.  They have diverse  applications
in social  networks,   biology,  and  economics \cite{gauss3,gauss5},
to  name  a  few.  Gaussian  trees  in  particular have attracted much
attention \cite{gauss5} due to their sparse structures, as well as
existing computationally efficient algorithms in learning the underlying
topologies  \cite{mit,correlation}.  

In a latent Gaussian tree, we are dealing with two sets of variables. 
Let $\mathbf{X}=\{X_1,X_2,...,X_n\}$ be the $n$ observed variables in a Gaussian tree, i.e., the covariance matrix $\Sigma_{\mathbf{x}}$ is given. The set of variables $\mathbf{Y}=\{Y_1,Y_2,...,Y_k\}$ are hidden to us and should be estimated.
Note that for $\Sigma_{\mathbf{x}}$ to induce a latent Gaussian tree, it needs to satisfy certain conditions shown in \cite{correlation}.
In fact, for any triplet $x_i,x_j,x_k\in \mathbf{X}$ and writing $\rho_{x_ix_j}$ to show the pairwise correlation we need to have $|\rho_{x_ix_j}|\geq |\rho_{x_ix_k}\rho_{x_jx_k}|$ and $\rho_{x_ix_j}\rho_{x_ix_k}\rho_{x_jx_k}>0$. Such constraints on the correlation space shown to be necessary and sufficient for a joint Gaussian distribution to characterize a latent Gaussian tree \cite{correlation}.

There are several works such as \cite{mit,nj} that have proposed efficient algorithms to infer the latent Gaussian tree parameters. In fact, Choi et al., proposed a new \textit{recursive grouping} (RG) algorithm along with its improved version, i.e., \textit{Chow-Liu} RG (CLRG) algorithm to recover a latent Gaussian tree that is both \textit{structural} and \textit{risk} consistent \cite{mit}, hence it recovers the \textit{correct} value for the latent parameters.
They introduced a \textit{tree metric} as the negative \textit{log} of the absolute value of pairwise correlations to perform the algorithm.

In this paper we assume that the
parameters and structure information of the latent Gaussian tree is
provided using one of aforementioned algorithms.

Our primary concern in such synthesis problem is about efficiency in terms of the amount of random bits required at the input, as well as the modeling complexity of given stochastic system through which the Gaussian vector is synthesized.
Such efficiency is characterized through defining proper random \textit{sequences}, and random \textit{bins} containing those sequences, which we define as random \textit{codewords} and \textit{codeboooks}.
We use the input code-rate to define the complexity of our synthesis systems, since minimizing such rates results in reducing the number of common random bits needed to generate the output statistics.
In particular, through showing the intrinsic \textit{sign singularity} in latent Gaussian trees, we have demonstrated that such ambiguity can further help us to reduce the synthesis rates for such Gaussian trees.
To clarify, we consider the following case study.
\subsection{Motivating Case Study}

\begin{figure} [h]
\centering 
\includegraphics[width=0.5\columnwidth]{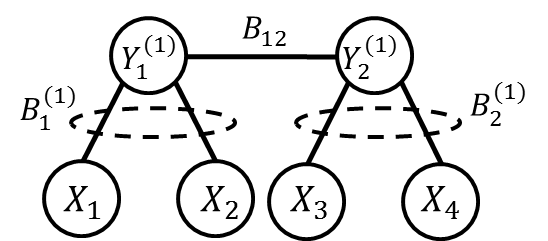}
\caption{A simple Gaussian tree with a hidden node $Y^{(1)}$\label{fig:broadcast}} 
\end{figure}

Consider a Gaussian tree shown in Figure \ref{fig:broadcast}.
It consists of four observed variables $X_1$, $X_2$, $X_3$, and $X_4$ that are connected to each other through two hidden nodes $Y^{(1)}_1$ and $Y^{(1)}_2$.
Define \small$\rho_{x_1y_1}=E[X_1Y_1^{(1)}]$ \normalsize as the true correlation value (edge-weight) between the input $Y_1^{(1)}$ and the output $X_1$.
We can similarly define other correlation values $\rho_{x_2y_1}$, $\rho_{x_3y_2}$, and $\rho_{x_4y_2}$.
Define $B^{(1)}_j\in\{-1,1\},~j\in[1,2]$ \normalsize as a binary variable corresponding to the $j-th$ input that as we will see reflects the sign information of pairwise correlations. 
For the tree shown in Figure \ref{fig:broadcast}, one may assume that $B^{(1)}_j=1$ to show the case with $\rho'_{x_iy_j}=\rho_{x_iy_j}$, while $B^{(1)}_j=-1$ captures $\rho''_{x_iy_j}=-\rho_{x_iy_j}$, where $\rho'_{x_iy_j}$ and $\rho''_{x_iy_j}$, $i\in[1,2]~or~i\in[3,4]$ are the (alternative) recovered correlation values using certain inference algorithm such as RG \cite{mit}.
Also, define $B_{12} = B^{(1)}_1B^{(1)}_2$.
It is easy to see that both recovered correlation values induce the same covariance matrix $\Sigma_{\mathbf{x}}$, showing the sign singularity issue in such a latent Gaussian tree.
In particular, for each pairwise correlation $\rho_{x_kx_l},k<l \in [1,2,3,4]$, and if $x_k$ and $x_l$ have the same parent, we have $\rho_{x_kx_l}=\rho_{x_ky_j}\rho_{x_ly_j}=(B^{(1)}_j)^2\rho_{x_ky_j}\rho_{x_ly_j}$, where the second equality is due to the fact that regardless of the sign value, the term $(B^{(1)}_j)^2$ is equal to $1$.
Now, depending on whether we replace $B^{(1)}_j$ with $\{1,-1\}$, we obtain $\rho_{x_kx_l}=\rho'_{x_ky_j}\rho'_{x_ly_j}=\rho''_{x_ky_j}\rho''_{x_ky_j}$.
And there is no way to distinguish these two groups using only the given information on observables joint distribution.
Similarly, if $x_k$ and $x_l$ are connected to different input nodes, we can write $\rho_{x_kx_l}=\rho_{x_ky_1}\rho_{x_ly_2}=B^{(1)}_1 B^{(1)}_2 B_{12}\rho_{x_ky_j}\rho_{x_ly_j}$, where the second equality is due to $B_{12} = B^{(1)}_1B^{(1)}_2$.
Again, one cannot recover the sign information from only the output correlation values.

Such sign singularity patterns become more complex as the tree size grows.
In section \ref{sec:main} we characterize certain properties of sign information.

It turns out that such sign singularity can be seen as another \textit{noisy} source of randomness, which can further help us to reduce the code-rate corresponding to latent inputs to synthesize the latent Gaussian tree.
In fact, we may think of the Gaussian tree shown in Figure \ref{fig:broadcast} as a communication channel, where information flows from the source \small $\mathbf{Y}^{(1)}=[Y^{(1)}_1,Y^{(2)}_2]$ \normalsize through four channels $p_{X_i|Y^{(1)}_j}$ with independent additive Gaussian noise variables \small $Z_i\sim N(0,\sigma^2_{z_i}),~i\in\{1,2,3,4\}$ \normalsize to generate (dependent) outputs with \small $\mathbf{X}\sim N(0,\Sigma_{\mathbf{x}})$. 
\normalsize We introduce \small $\mathbf{B}^{(1)}=[B^{(1)}_1,B^{(2)}_2]\in\{-1,1\}$ \normalsize as binary Bernoulli random variables with parameters $\pi_{\mathbf{B}^{(1)}}$ and $\pi_{\mathbf{B}^{(2)}}$, which reflect the sign information of pairwise correlations.
In fact, we may define the following affine transformation from inputs to outputs,
\begin{align} \label{eq:dumbbell_affine}
\begin{bmatrix}
X_1\\
X_2\\
X_3\\
X_4
\end{bmatrix}
=
\begin{bmatrix}
\alpha_{11}B^{(1)}_1 & 0\\
\alpha_{21}B^{(1)}_1 & 0\\
0 & \alpha_{32}B^{(1)}_2\\
0 & \alpha_{42}B^{(1)}_2
\end{bmatrix}
\begin{bmatrix}
Y^{(1)}_1\\
Y^{(1)}_2
\end{bmatrix}
+
\begin{bmatrix}
Z_1\\
Z_2\\
Z_3\\
Z_4
\end{bmatrix}
\end{align}

\noindent where $\alpha_{ij}$ are given values that characterize the correlations up to sign, i.e., $|\rho_{x_iy_j}|$.

Our goal is to characterize the achievable rate region and through a synthesis scheme to generate Gaussian trees with density $q_{\mathbf{X}\mathbf{Y}^{(1)}}$ using only the hidden inputs and through a channel with additive Gaussian noises, where the synthesized joint density $q_{\mathbf{X}\mathbf{Y}^{(1)}}$ is indistinguishable from the true Gaussian tree density $p_{\mathbf{X}\mathbf{Y}^{(1)}}$ as measured by \textit{total variation} metric \cite{cuff}.
To achieve this, we first generate many sampled sequences (codewords) $(\mathbf{Y}^{(1)})^N$ and $(\mathbf{B}^{(1)})^N$ to form the corresponding bins (codebooks) containing those codewords.
The size of the codebooks are characterized by $2^{NR_{\mathbf{Y}^{(1)}}}$ and $2^{NR_{\mathbf{B}^{(1)}}}$, where $R_{\mathbf{Y}^{(1)}}$ and $R_{\mathbf{B}^{(1)}}$ are the codebook rates, regarding to sign and hidden nodes codewords.
Each time to generate output sequences, we first randomly pick sign and latent node codewords and then we use the synthesis channel in \eqref{eq:dumbbell_affine} to achieve a particular output sequence $\mathbf{X}^N$.
We aim to characterize the lower bound on the codebook rates, through which the generated sequence's statistics, i.e., $q_{\mathbf{X}\mathbf{Y}^{(1)}}$ is asymptotically (as $N\rightarrow \infty$) indistinguishable from the desired statistics.
In particular, we characterize the quantity $\inf_{p_{\tilde{\mathbf{Y}}^{(1)}}} I(\mathbf{X};\tilde{\mathbf{Y}}^{(1)})$, where $I(\mathbf{X};\tilde{\mathbf{Y}}^{(1)})$ is the mutual information between the output $\mathbf{X}$ and the input vector $\tilde{\mathbf{Y}}^{(1)}=\{\mathbf{Y}^{(1)},\mathbf{B}\}$.
This corresponds to finding the minimum achievable rate under Gaussian tree assumption.
Equivalently, we are seeking for optimal values of $\pi_{\mathbf{B}^{(1)}}$ and $\pi_{\mathbf{B}^{(2)}}$ to maximize the achievable rate region characterized by $R_{\mathbf{Y}^{(1)}}$ and $R_{\mathbf{B}^{(1)}}$.
\begin{rmk}
Suppose for a moment, instead of using the tree structure, we simply used six independent normalized Gaussian variables and by passing them through a filter, i.e., linear combination of these independent variables, we produce the desired Gaussian tree (with two hidden nodes and fours observables).
While this approach seems appealing, note that as it is observed in \cite{steeg} as well, such synthesis scheme needs infinite bits of precision to produce the desired statistics, which is practically infeasible.
This is due to the noiseless nature of the channel (see \textit{channel resolvability} \cite{verdu}), i.e., the linear filter, which is noise-free that makes the input code rates maximized (since the input-output mutual information will be maximized), hence, we need infinite bits of precision to synthesize the desired Gaussian density.
In contrast, our framework exploits the tree structure to further reduce the rates needed for synthesis.
Moreover, to characterize the channel shown in Figure \ref{fig:broadcast}, one may need to introduce only four parameters $\alpha_{ij}$, one for each edge, while the aforementioned \textit{naive} approach needs nine parameters (basically each input is connected to all the outputs) to capture the dependency structure of output variables.
This modeling efficiency will become more evident in more general and larger Gaussian trees, since in that case the naive approach faces with $\mathcal{O}(n^2)$ parameters while our approach only needs $\mathcal{O}(n+k)$ parameters (in the order of edge-set cardinality of the tree), where $n$ and $k$ are the number of outputs, and latent inputs, respectively.
Such efficiency is captured by sparsity structure of connection matrix $A_B$ between the input and output, which will be completely characterized in subsequent sections.
\end{rmk}

\subsection{Related Works}
Wyner's Common information characterizes the minimum amount of common randomness needed to approximate the joint density between a pair of random variables $X_1$ and $X_2$ to be $C(X_1,X_2)=\min_{\substack{P_Y\\X_1-Y-X_2}} I(X_1,X_2;Y)$, where $C(X_1,X_2)$ is widely known as \textit{Wyner's common information}.
This is done through a common source of randomness, i.e., $Y$, and two independent random sources to generate $X_1$ and $X_2$ with desired joint statistics.
Han and Verdu, in \cite{verdu} along the same problem, define the notion of \textit{resolvability} of a given channel, which is defined as the minimal required randomness to generate output statistics in terms of a vanishing total variation distance between the synthesized and prescribed joint densities.
Resolvability of a channel is found to be a very intuitive description of common randomness in our settings, since it can be related to channel quality in terms of its noise power, and the noisier the channel the less number of common random bits needed to simulate the output \cite{verdu}.
Along the same line, Cuff in \cite{cuff} completely characterized the achievable rate regions needed to synthesize a memoryless channel, where he also used the total variation distance metric to show the quality of the proposed scheme.

There are several works that extend the classical bi-variate synthesis problem in Wyner's study to more general scenarios.
In \cite{allerton2011,isit2014,trans2016}, the authors aim to define the common information of $n$ dependent random variables, to further address the same question in this setting.
A lower bound on such generalized common information is obtained in \cite{lower_wyner}.
Also, the common information for a special case with $n$ Gaussian variables with homogeneous pairwise correlations is obtained.
They resort to the same scenario as Wyner \cite{wyner} did, i.e., considering one random variable to define such common randomness.
Veld and Gastpar \cite{caching1} characterize such quantity for a more general set of Gaussian vectors with circulant covariance matrices.
Also, in \cite{cuff_common} the authors completely characterize the common information between two jointly Gaussian vectors, as a function of certain singular values that are related to both joint and marginal covariance matrices of two Gaussian random vectors.
However, they still divide the random vector into two groups, which makes it similar to Wyner's scenario.

In this paper, we are not concerned with solving the common information problem for Gaussian trees.
Instead, we want to motivate the notion \textit{multi-variable} synthesis, that is instead of introducing a single variable $Y$, we define a random vector $\mathbf{Y}$ with certain dependency structure to capture the common randomness and produce common random bits.
We provide a layered synthesis algorithm, along with the corresponding achievability regions to synthesize those distributions inducing a Gaussian tree.
In \cite{gastpar} such general case is appropriately defined using a constrained convex optimization problem.
The benefits of such general assumption is shown in \cite{steeg}.
In fact, Steeg \textit{et. al.} implement a new method based on \textit{multi-layer} common variables for a particular blind source separation problem and showed that their proposed model outperforms all previous learning algorithms.

Similar to \cite{steeg,gastpar} we also consider multi-variable cases, but unlike those works, we are interested in characterizing the achievable rates to synthesize a special class of Gaussian distributions, namely Gaussian trees.
We adopt a specific (but natural) structure
to our synthesis scheme to decrease the number parameters to model the
synthesis scheme. 
It is worthy to point that the achievability results
given in this paper are under the assumed structured synthesis framework.
Hence, although through defining an optimization problems, we show that
the proposed method is efficient in terms of both modeling and codebook
rates, the converse proof, which shows the optimality of such scheme
and rate regions is never claimed.

\subsection{Contributions}
Our main contributions can be summarized as follows:


$\bullet$ We propose a novel generative modeling scheme, by which we synthesize any Gaussian vector that lies in a subspace of latent Gaussian trees.
The proposed scheme is modeling-wise efficient, since by relying on the inferred latent tree structure it reduces the number of parameters needed at each step for output synthesis. 
We also characterize the achievable rate regions for all the channels at each layer.


$\bullet$ We prove that under the latent Gaussian tree assumption, the mutual information between the output vector and both latent inputs and sign variables is only a function of output's covariance matrix $\Sigma_{\mathbf{X}}$.
We provide a general formula for such mutual information in a case of \textit{leaf} outputs.
We also show that given the sign information, the mutual information between each adjacent layer vectors is fixed as well.
We show that the achievable rates are lower bounded by input-output mutual information values at each layer.

$\bullet$ We show that the lower bounds on latent variable rates are a function of Bernoulli sign variables.
Such sign ambiguity can be seen as another source of randomness to further help us achieve lower codebook rates for synthesis.
We prove that such lower bounds can be minimized (hence maximizing the achievable rate region) in a case of homogeneous Bernoulli distributed sign information.

$\bullet$ In our previous work \cite{arxiv_version}, we only characterized the achievable rate regions for output synthesis of the latent Gaussian trees with leaf observables, and with each hidden node only connected to the upper layer inputs. 
However, in this paper not only we provide a constructive proof for those subclass of Gaussian trees, but also we completely characterize the synthesis scheme to generate the entire statistics of any latent Gaussian tree structure.

\subsection{Paper Organization}
The rest of the paper is organized as follows. Section \ref{sec:formulation}
gives the problem formulation and models the sign singularity in latent Gaussian trees.
In section \ref{sec:main} we first show a direct relation between the number of latent inputs and the needed sign input variables.
Then, for any general Gaussian tree we prove that the input-output mutual information at each layer is only a function of output statistics.
In section \ref{sec:achievable} we address the problem of layered synthesis of Gaussian trees through three different case studies, which cover all possible situations that may happen in such structures.
General bounds on achievable rate tuples is characterized for all scenarios.
We conclude the paper in Section \ref{sec:conclusion}.

\section{Problem Formulation} \label{sec:formulation}

\subsection{The signal model of a multi-layer latent Gaussian tree}
Here, we suppose a latent graphical model, with $\mathbf{Y}=[Y_1,Y_2,...,Y_k]'$ as the set of inputs (hidden variables), $\mathbf{B}=[B_1,...,B_m]$, with each $B_i\in\{-1,1\}$ being a binary Bernoulli random variable with parameter $\pi_i=p(B_i=1)$ to introduce sign variables, and $\mathbf{X}=[X_1,X_2,...,X_n]'$ as the set of Gaussian outputs (observed variables) with $p_{\mathbf{X}}(\mathbf{x})$.
We also assume that the underlying network structure is a latent Gaussian tree, therefore, making the joint probability (under each sign realization) $p_{\mathbf{XY|B}}$ be a Gaussian joint density $N(\mathbf{\mu},\Sigma_{\mathbf{xy|b}})$, where the covariance matrix $\Sigma_{\mathbf{xy|b}}$ induces tree structure $G_T(V,E,W)$, where $V$ is the set of nodes consisting of both vectors $\mathbf{X}$ and $\mathbf{Y}$; $E$ is the set of edges; and $W$ is the set of edge-weights determining the pairwise covariances between any adjacent nodes.
We consider normalized variances for all variables $X_i\in \mathbf{X},~i\in\{1,2,...,n\}$ and $Y_j\in\mathbf{Y},~j\in\{1,2,...,k\}$. Such constraints do not affect the tree structure, and hence the independence relations captured by $\Sigma_{\mathbf{xy|b}}$.
Without loss of generality, we also assume $\mathbf{\mu}=\mathbf{0}$, this constraint does not change the amount of information carried by the observed vector. 

In \cite{arxiv_version} we showed that the vectors $\mathbf{X}$ and $\mathbf{B}$ are independent, and the intrinsic sign singularity in Gaussian trees is due to the fact that the pairwise correlations $\rho_{x_ix_j}\in \Sigma_\mathbf{x}$ can be written as $\prod_{(l,k)\in E} \rho_{x_lx_k}$, i.e., the product of correlations on the path from $x_i$ to $x_j$.
Hence, roughly speaking, one can carefully change the sign of several correlations of the path, and still maintain the same value for $\rho_{x_ix_j}$.
Although this results in no variation on the correlation values $\rho_{x_nx_m},~n,m\in V$, we showed that if the cardinality of the input vector $\mathbf{Y}$ is $k$, then $2^k$ minimal Gaussian trees (that only differ in sign of pairwise correlations) may induce the same joint Gaussian density $p_{\mathbf{X}}$ \cite{arxiv_version}.

In order to propose the successive synthesis scheme, we need to characterize the definition of \textit{layers} in a latent Gaussian tree.
We define latent vector $\mathbf{Y}^{(l)}$, to be at layer $l$, if the shortest path between each latent input $Y_i^{(l)}\in\mathbf{Y}^{(l)}$ and the observed layer (consisting the output vector $\mathbf{X}$) is through $l$ edges.
In other words, beginning from a given latent Gaussian tree, we assume the output to be at layer $l=0$, then we find its immediate latent inputs and define $\mathbf{Y}^{(1)}$ to include all of them.
We iterate such procedure till we include all the latent nodes up to layer $L$, i.e., the top layer.
In such setting, the sign input vector $\mathbf{B}^{(l)}$ with Bernoulli sign random variables $B_i^{(l)}\in\mathbf{B}^{(l)}$ is assigned to the latent inputs $\mathbf{Y}^{(l)}$.

We adopt a synthesis channel to feature the relationship between each pair of successive layers.
Assume $\mathbf{Y}^{(l+1)}$ and $\mathbf{B}^{(l+1)}$ as the input vectors, $\mathbf{Y}^{(l)}$ as the output vector, and the noisy channel to be characterized by the conditional probability distribution $P_{\mathbf{Y}^{(l)}|\mathbf{Y}^{(l+1)},\mathbf{B}^{(l+1)}}(\mathbf{y}^{(l)}|\mathbf{y}^{(l+1)},\mathbf{b}^{(l+1)})$, the signal model for such a channel can be written as follows,
\begin{align} \label{eq:multi-layer_linear_regression}
\mathbf{Y}^{(l)}=\mathbf{A_B}^{(l,l+1)}\mathbf{Y}^{(l+1)}+\mathbf{Z}^{(l+1)},
~~~l\in[0,L-1]
\end{align}
\noindent \noindent where $\mathbf{Z}^{(l+1)}\sim N(0,\Sigma_{\mathbf{z}^{(l+1)}})$ is the additive Gaussian noise vector with independent elements, each corresponding to a different edge from the input layer $l+1$ to the output layer $l$.
Also, $\mathbf{A_B}^{(l,l+1)}$ is the $|\mathbf{Y}^{(l)}|\times |\mathbf{Y}^{(l+1)}|$ sparse connection matrix that also carries the sign information vectors $\mathbf{B}^{(l+1)}$ and $\mathbf{B}^{(l+1)}$. 
The sparsity of the transition matrix $\mathbf{A_B}^{(l,l+1)} = [\alpha_{ij}]$ is due to assumed underlying tree structure, and it follows the following form
\begin{align}
\alpha_{ij}
=
  \begin{cases}
  \gamma_{ij}b^{(l)}_i b^{(l+1)}_j & e_{ij}\in E\\
  0 & e_{ij}\notin E\\
  \end{cases}
\end{align}
\noindent where $e_{ij}$ denotes the edge between $Y_{i}^{(l)}$ and $Y_{j}^{(l+1)}$.
The existence of such edge can be verified from the set $E$, which is obtained during the learning process.
Also, $\gamma_{ij}$ is the edge-weight showing the correlation value $\rho_{ij}$ up to a sign.
Note that, the case for $l=0$ is a special case, where $\mathbf{A_B}^{(0,1)}$ only depends on $\mathbf{B}^{(1)}$ since there is no sign singularity at the observable layer.

The outputs $\mathbf{Y}^{(l)}$ at each layer $l$, are generated using the inputs $\mathbf{Y}^{(l+1)}$ at the upper layer.
As we will see next, such modeling will be the basis for our successive synthesis scheme.
In fact, by starting from the top layer inputs $L$, at each step we generate the outputs at the lower layer, this will be done till we reach the observed layer to synthesize the Gaussian vector $\mathbf{X}$.
Finally, note that in order to take all possible latent tree structures, we need to revise the ordering of layers in certain situations, which will be taken care of in the following subsections.
For now, the basic definition for layers will be satisfactory.

\subsection{Synthesis Approach Formulation}
Here, we provide mathematical formulations to address the following fundamental problem: using channel inputs $\mathbf{Y}^{(l+1)}$ and $\mathbf{B}^{(l+1)}$, what are the rate conditions under which we can synthesize the Gaussian channel output $\mathbf{Y}^{(l)}$ with a given $p_{\mathbf{Y}^{(l)}|\mathbf{B}^{(l)}}$.
The synthesis channel at each layer is characterized by \eqref{eq:multi-layer_linear_regression}, where the random sequences at any lower layer are affine transformations of their corresponding random upper layer sequences.
Note that, at first we are only given $p_{\mathbf{X}}$, but using certain tree learning algorithms we can find those jointly Gaussian latent variables $p_{\mathbf{Y}^{(l)}|\mathbf{B}^{(l)}}$ at every level $l\in [1,L]$. 
In fact, to account for sign ambiguity we have to deal with mixture Gaussian vectors $p_{\mathbf{Y}^{(l)}}$ at each layer $l$.
We propose a successive synthesis scheme on multiple layers that together induce a latent Gaussian tree, as well as the corresponding bounds on achievable rate tuples.
The synthesis scheme is efficient because it utilizes the latent Gaussian tree structure to synthesize the output at each layer.
In particular, without resorting to such learned structure we need to characterize $\mathcal{O}(kn)$ parameters (one for each link between a latent and output node) in total, while by considering the sparsity reflected in a tree and each of transition matrices $\mathbf{A_B}^{(l,l+1)}$ we only need to consider $\mathcal{O}(k+n-1)$ parameters (the edges of a tree).

Suppose we \textit{transmit} input messages through $N$ channel uses, in which $t\in\{1,2,...,N\}$ denotes the time index.
\textit{Transmitting} a random sequence at each layer is equivalent to compute its mapping (the output sequence) through a synthesis channel defined in \eqref{eq:multi-layer_linear_regression}.
We define $\vec{Y}^{(l)}_{t}[i]$ to be the $t$-th symbol of the $i$-th codeword, with $i\in C_{\mathbf{Y}^{(l)}}=\{1,2,...,M_{Y^{(l)}}\}$ where $M_{Y^{(l)}}=2^{NR_{Y^{(l)}}}$ is the codebook cardinality, transmitted from the existing $k_l$ sources at layer $l$.
We assume there are $k_l$ sources $Y^{(l)}_j$ present at the $l$-th layer, and the channel has $L$ layers.
We can similarly define $\vec{B}^{(l)}_{t}[k]$ to be the $t$-th symbol of the $k$-th codeword, with $k\in C_{\mathbf{B}^{(l)}}=\{1,2,...,M_{B^{(l)}}\}$ where $M_{B^{(l)}}=2^{NR_{B^{(l)}}}$ is the codebook cardinality, regarding the sign variables at layer $l$.
We will further explain that although we define codewords for the Bernoulli sign vectors as well, they are not in fact transmitted through the channel, and rather act as \textit{noisy sources} to select a particular sign setting for latent vector distributions.
For \textit{sufficiently} large rates $R_{\mathbf{Y}}=[R_{Y^{(1)}},R_{Y^{(2)}},...,R_{Y^{(L)}}]$ and $R_{\mathbf{B}}=[R_{B^{(1)}},R_{B^{(2)}},...,R_{B^{(L)}}]$ and as $N$ grows the synthesized density of latent Gaussian tree converges to $p_{\mathbf{W}^N(\mathbf{w}^N)}$, i.e., $N$ i.i.d realization of the given output density $p_{\mathbf{W}}(\mathbf{w})$, where $W=\{\mathbf{X,Y,B}\}$ is a compound random variable consisting the output, latent, and sign variables.
In other words, the average total variation between the two joint densities vanishes as $N$ grows \cite{cuff},
\begin{align} \label{eq:TV}
\lim_{N\rightarrow\infty} E||q(\mathbf{w}_1,...,\mathbf{w}_N)-\prod_{t=1}^N p_{\mathbf{w}_t}(\mathbf{w}_t)||_{TV}\rightarrow 0
\end{align}
\noindent where $q(\mathbf{w}_1,...,\mathbf{w}_N)$ is the synthesized density of latent Gaussian tree, and $E||.||_{TV}$, represents the average total variation.
In this situation, we say that the rates $(R_{\mathbf{Y}},R_{\mathbf{B}})$ are \textit{achievable} \cite{cuff}.
Our achievability proofs heavily relies on \textit{soft covering lemma} shown in \cite{cuff}.
Loosely speaking, the soft covering lemma states that one can synthesize the desired statistics with arbitrary accuracy, if the codebook sizes (or equivalently, the rates $(R_{\mathbf{Y}},R_{\mathbf{B}})$) are sufficient and the channel through which these codewords are sent is noisy enough.
This way, one can cover the desired statistics up to arbitrary accuracy.
The main objective is to maximize such rate region, and develop a proper synthesis scheme to achieve the desired statistics.

For simplicity of notation, we drop
the symbol index and use $Y^{(l)}_{t}$ and $B^{(l)}_{t}$ instead of
$\vec{Y}^{(l)}_{t}[i]$ and $\vec{B}^{(l)}_{t}[k]$, respectively, since
they can be understood from the context.

\section{Mutual Information of Layered Synthesis Channels with Correlation Sign Singularity} \label{sec:main} 
\subsection{Properties of sign information vector $\mathbf{B}$}

In Theorem \ref{thm:B}, whose proof can be found in Appendix \ref{app:B}, we characterize the size and dependency relations of sign vectors for any general minimal latent Gaussian tree.

\begin{thm} \label{thm:B}
{\it
$(1)$ The correlation values $\rho_{yx_i}$ in regard to the outputs $X_i$ that are connected to a single input, say $Y$, share an equivalent sign class, i.e., they either all belong to $B=b$ or $B=-b$.

$(2)$ Given the cardinality of input vector $\mathbf{Y}=\{Y_1,Y_2,...,Y_k\}$ is $k$, then there are totally $2^k$ minimal Gaussian trees with isomorphic structures, but with different correlation signs that induce the same joint density of the outputs, i.e., equal $p_{\mathbf{X}}(\mathbf{x})$.
}
\end{thm}

For example, in a Gaussian tree shown in Figure \ref{fig:broadcast}, there is only one hidden node $Y^{(1)}$, and we already know by previous discussions that there are two latent Gaussian trees with different sign values for $B^{(1)}$, which induce the same output joint density $p_{\mathbf{X}}(\mathbf{x})$.
In more general cases the problem of assigning correlation sign variables is more subtle, where we clarify the approach using two examples, next.
\begin{figure}[h!]
    \begin{subfigure}[h]{\columnwidth}
    \centering
        \includegraphics[width=0.35\columnwidth]{B1}
        \caption{}
        \label{fig:B1}
    \end{subfigure}
    
    ~ 
    \begin{subfigure}[h]{\columnwidth}
    \centering
        \includegraphics[width=0.6\columnwidth]{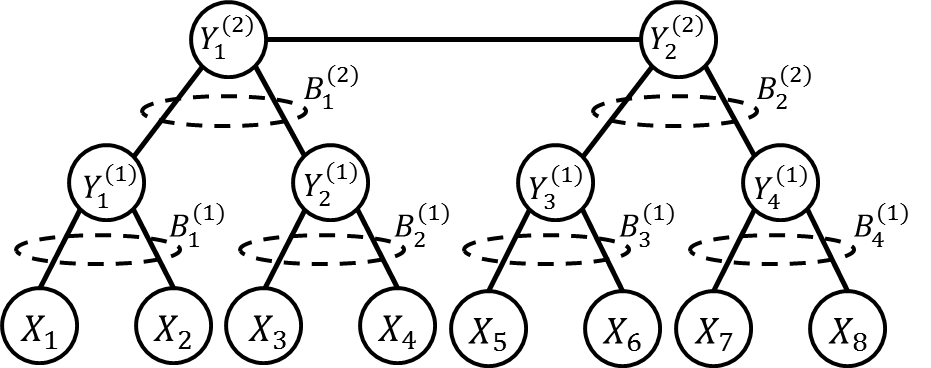}
        \caption{}
        \label{fig:B2}
    \end{subfigure}
    \caption{Two possible cases to demonstrate the dependency relations of sign variables: (a) with two hidden inputs, and (b) with $4$ hidden inputs at two layers}\label{fig:B}
\end{figure}

In a Gaussian tree shown in Figure \ref{fig:B}(a) there are two hidden nodes $Y_1$ and $Y_2$. By Theorem \ref{thm:B}, we know that there are $4$ Gaussian trees with sign ambiguity. 
Also, from the first part in Theorem \ref{thm:B} we may introduce $B^{(1)}_1$ to capture the correlation signs $\rho_{x_1y_1}$ and $\rho_{x_2y_1}$, and $B^{(1)}_2$ for the correlation signs $\rho_{x_3y_2}$ and $\rho_{x_4y_2}$.
We can think of $B_{12}$ as the sign of $\rho_{y_1y_2}$.
Note that the link between the variables $Y_1$ and $Y_2$ are in both groups with common correlation sign, so we anticipate that $B_{12}$ should be dependent on both $B^{(1)}_1$ and $B^{(1)}_2$.
Since we need to maintain the correlation signs regarding $\rho_{x_ix_j},~i\in\{1,2\},~j\in\{3,4\}$, hence the product $B^{(1)}_1B^{(1)}_2B_{12}$ should maintain its sign. 
Thus, we have $B_{12}=B^{(1)}_1B^{(1)}_2$, so $B_{12}$ is completely determined given $B^{(1)}_1$ and $B^{(1)}_2$. 
In other words, we may write the pairwise correlation as $E[Y_1^{(1)}Y_2^{(1)}]=\gamma_{12}B^{(1)}_1B^{(1)}_2$, which further justifies the Gaussian mixture assumption for latent variables $Y_1^{(1)}$ and $Y_1^{(2)}$.
Next, consider the Gaussian tree shown in Figure \ref{fig:B}(b), in which there are six hidden inputs. 
Similar to the previous case, we introduce four sign variables to capture the first layer sign information.
In this case, we further need to introduce $B^{(2)}_1$ and $B^{(2)}_2$ corresponding to second layer latent inputs.
Similar to the previous cases, the pairwise correlation sign of latent inputs depends on the corresponding sign variables at the same layer.
For example, $E[Y_1^{(1)}Y_4^{(1)}]=\gamma_{14}B^{(1)}_1B^{(1)}_4$ or $E[Y_1^{(2)}Y_2^{(2)}]=\gamma_{12}B^{(2)}_1B^{(2)}_2$.

One may note the pattern on such definition: The pairwise correlation signs of latent inputs only depends on the corresponding sign variables at the same layer, and it is independent of sign variables at other layers.
As we will see shortly, such property is essential in our layered synthesis scheme to make sure the conditional independence of different layers from each other, given the information of neighboring layers.

\subsection{Single Layer case: Mutual Information between Observables and Latent Variables}
It is best to start the achievability discussion by a simple scenario we considered in \cite{arxiv_version}.
In \cite{arxiv_version} we were only concerned about the synthesis of output vector statistics with given $p_{\mathbf{X}}$.
Let us define $\mathbf{\tilde{Y}}=\{\mathbf{Y},\mathbf{B}\}$, then the formalized problem has the following form:
\begin{align} \label{eq:common_info}
\inf&_{p_\mathbf{\tilde{Y}}(\mathbf{\tilde{y}})} I(\mathbf{X};\mathbf{\tilde{Y}}),~s.t.,\notag\\
&p_{\mathbf{X},\mathbf{\tilde{Y}}}(\mathbf{x},\mathbf{\tilde{y}})~induces~a~minimal~Gaussian~tree\notag\\
&X_i\perp X_j|\mathbf{\tilde{Y}}\notag\\
&\Sigma_{\tilde{y}\in\mathbf{\tilde{Y}}} p(\mathbf{x},\mathbf{\tilde{y}})=p_{\mathbf{X}}(\mathbf{x})
\end{align}

\begin{rmk}
Due to Markov property, we know that given $\tilde{Y}^{(1)}$, the output layer $\mathbf{X}$ is conditionally independent of all vectors $\tilde{Y}^{(l)},~l\in [2,L]$ at upper layers.
Hence, we have the equality $I(\mathbf{X};\mathbf{\tilde{Y}}) = I(\mathbf{X};\mathbf{\tilde{Y}}^{(1)})$, i.e., to synthesize the output vector statistics, all we need are the common latent inputs $\tilde{Y}^{(1)}$ (and of course the independent additive Gaussian noises and Bernoulli sign variables).
As we will see shortly, this is a special case to our layered synthesis strategy, where we only deal with a single layer, and want to synthesize the output statistics.
\end{rmk}
\begin{rmk}
Note that such optimization problem is defined for those output vectors $\mathbf{X}$, whose covariance matrix $\Sigma_{\mathbf{X}}$ is in the subspace of positive definite matrices that induce a latent Gaussian tree.
As discussed earlier, such subspace can be completely characterized by a systems certain inequalities (or equalities in certain cases) between pairwise covariance elements in $\Sigma_{\mathbf{X}}$ \cite{correlation}.
Hence, all of the mutual information values should be evaluated under a given Gaussian tree $G_T(V,E,W)$. 
For simplicity we drop this notation in their expressions.
Hence, such problem is not the same as the general \textit{Wyner's common information} setting, since in Wyner's scenario, no structural constraint is imposed on latent variables.
\end{rmk}

The \textit{minimality} assumption on the Gaussian tree structure, indicates that in our case $|\mathbf{X}|\geq 3$, i.e., the number of observed variables should be at least three.
In a minimal Gaussian tree we assume all the hidden variables have at least three neighbors \cite{mit}, which results in ignoring all those singular cases where there can be arbitrarily redundant hidden variables added to the model without changing the observed joint density $p_{\mathbf{X}}(\mathbf{x})$.
In this setting, by Theorem \ref{thm:fixed}, whose proof can be found in Appendix \ref{app:fixed}, we show that  regardless of the underlying Gaussian tree structure, there is no room to minimize $I(\mathbf{X};\tilde{\mathbf{Y}})$.

\begin{thm} \label{thm:fixed}
{\it
Given $p_{\mathbf{X}}(x)\sim~N(0,\Sigma_\mathbf{x})$ and the settings in \eqref{eq:common_info}, the mutual information $I(\mathbf{X};\mathbf{\tilde{Y}})$ is only a function of $\Sigma_\mathbf{x}$ and if the observable nodes are only leaf nodes, the mutual information is given by,
\begin{align} \label{eq:fixed_mutual_information}
I(\mathbf{X};\mathbf{\tilde{Y}})=\dfrac{1}{2}\log\dfrac{|\Sigma_\mathbf{x}|}{\prod_{i=1}^n (1-\dfrac{\rho_{x_ix_{j_i}}\rho_{x_ix_{k_i}}}{\rho_{x_{j_i}x_{k_i}}})}
\end{align}
\noindent where for each $X_i$, we choose two other nodes $X_{j_i}$, $X_{k_i}$, where all three of them are connected to each other through $Y_{X_i}$ (i.e., one of their common ancestors), which is one of the hidden variables adjacent to $X_i$. 
}
\end{thm}

Intuitively, given $\Sigma_x$ and any three outputs that have a common latent variable as their input, the correlation values between each output and the input is fixed, since varying one correlation results in varying the other correlations in the same direction, hence making the pairwise correlation between the other outputs change, which is impossible. 
\begin{rmk} \label{rmk:fixed-edge-weights}
Theorem \ref{thm:fixed} indicates a special behavior of the mutual information under latent Gaussian tree assumption.
In particular, given $X_i$ and its latent parent $Y_{X_i}$ we may end up with several options for $X_{j_i}$ and $X_{k_i}$.
However, it can be shown that in a subspace of correlations corresponding to latent Gaussian trees \cite{correlation}, all those distinct options result in a same value for the term $\rho_{x_ix_{j_i}}\rho_{x_ix_{k_i}}/\rho_{x_{j_i}x_{k_i}}$.
In fact, we show that such terms are all equal to $\rho^2_{x_iy_{x_i}}\in (0,1)$, for $X_i\in\mathbf{X}$ and $Y_{X_i}\in\mathbf{Y}^{(1)}$.
In other words, they characterize the correlation between each of the outputs $X_i$ with its corresponding parent $Y_{X_i}$ at the first layer.
\end{rmk}
\begin{rmk}
Due to equality $I(\mathbf{X};\mathbf{Y},\mathbf{B})=I(\mathbf{X};\mathbf{Y}^{(1)},\mathbf{B}^{(1)})$ we can show that to compute the mutual information value in Theorem \ref{thm:fixed}, we only need the correlation values of them form $\rho^2_{x_iy_{pa_i}}$ that are between the observables and their immediate parents.
As we will see shortly, this argument can be easily generalized to a multi-layer case, in which to compute the mutual information between the outputs of each layer and the higher layer variables, we only need those inputs that are the parents of output variables, i.e., the variables in a single layer above the outputs.
\end{rmk}

Note that from \eqref{eq:fixed_mutual_information} we can see that the mutual information $I(\mathbf{X};\mathbf{Y,B})$ does not depend on sign information, which further justifies our point on intrinsic sign ambiguity in latent Gaussian trees.
One may easily deduce the following,
\begin{align} \label{eq:equality}
I(\mathbf{X};\mathbf{\tilde{Y}})=I(\mathbf{X};\mathbf{Y},\mathbf{B})=I(\mathbf{X};\mathbf{Y}) + I(\mathbf{X};\mathbf{B}|\mathbf{Y})
\end{align}
The results in Theorem \ref{thm:fixed} combined with \eqref{eq:equality}, suggests that by minimizing $I(\mathbf{X};\mathbf{Y})$, one may eventually maximize $I(\mathbf{X};\mathbf{B}|\mathbf{Y})$ 
, i.e., quantifying the maximum amount of information loss on the sign input $\mathbf{B}$.
In other words, to reach lower synthesis rates and maximizing the achievable rate region, we need to maximize the information loss on sign information.
In Theorem \ref{thm:uniform_sign}, whose proof can be found in Appendix \ref{app:uniform_sign} we show that in order to minimize the mutual information $I(\mathbf{X};\mathbf{Y})$ the sign inputs should be uniformly distributed.
\begin{thm} \label{thm:uniform_sign}
{\it
Given the Gaussian vector $\mathbf{X}$ with $\Sigma_{\mathbf{x}}$ inducing a latent Gaussian tree, with latent parameters $\mathbf{Y}$ and sign vector $\mathbf{B}$ the optimal solution for $\mathbf{\pi}^*=arg\min_{\mathbf{\pi}\in [0,1]^k} I(\mathbf{X};\mathbf{Y})$ happens for uniform sign vector.
}
\end{thm}

In other words, for all Bernoulli variables $B_i\in\mathbf{B}$ for the optimal solution we should have $\pi_i=1/2$.
Proving this result, relies upon showing the convexity of mutual information $I(\mathbf{X};\mathbf{Y})$ with respect to certain injective functions of $\pi_i$. Then, we show that the minimum happens for the case where all such functions are equal, and by converting these values back to $\mathbf{\pi}$-space we have the desired results.

\subsection{Multi-layer case: Mutual Information between outputs and inputs} \label{sec:multi_fixed}
Again, considering the successive synthesis perspective, we are interested in generating the output vector $\mathbf{Y}^{(l)}$, using its upper layer inputs $\mathbf{Y}^{(l+1)}$ with minimum amount of necessary random bits.
However, note that as shown in previous examples, in general $\mathbf{Y}^{(l)}$ follows a mixture Gaussian model, since its covariance matrix is dependent to the sign vector $\mathbf{B}^{(l)}$.
Hence, in order to follow the same principles as in \eqref{eq:common_info} the adopted objective function is $\inf_{p_{\mathbf{Y}^{(l+1)}}} I(\tilde{\mathbf{Y}}^{(l+1)};\mathbf{Y}^{(l)}|\mathbf{B}^{(l)})$, where conditioning on each realization of $\mathbf{B}^{(l)}$ results in a Gaussian density for the output vector $\mathbf{Y}^{(l)}$.
This is further discussed in detail, when we explain our successive synthesis method in the next subsection.
One may wonder whether the conditional independence and minimality constraints in \eqref{eq:common_info} also hold in this case.
The way we defined each input-output relation in \eqref{eq:multi-layer_linear_regression}, we can use similar arguments as before to show the independence of each output vector $\mathbf{Y}^{(l)}|\mathbf{B}^{(l)}$ with the sign input vector $\mathbf{B}^{(l+1)}$, since regardless of the sign input values the conditional output vectors remain jointly Gaussian.
Also, by the results of Theorem \ref{thm:fixed} we know that the overall mutual information $I(\mathbf{X};\tilde{\mathbf{Y}})$ is only a function of observed covariance matrix $\Sigma_\mathbf{x}$. So we may conclude that all the pairwise correlations in between any two consecutive layer are fixed, given $\Sigma_\mathbf{x}$. Intuitively,  such correlations are deduced from a latent tree, whose edge-weights are already determined via $\Sigma_\mathbf{x}$ (up to sign).
Hence, given $p_{\mathbf{X}}(x)\sim~N(0,\Sigma_\mathbf{x})$ and assuming $p_{\mathbf{X}\tilde{\mathbf{Y}}}$ induces a minimal latent Gaussian tree, the input-output mutual information at layers $l+1$ and $l$, i.e.,  $I(\tilde{\mathbf{Y}}^{(l+1)};\mathbf{Y}^{(l)}|\mathbf{B}^{(l)})$ for $l\in[0,L-1]$ is already determined by $\Sigma_\mathbf{x}$.

\section{Achievable Rate Regions for Successive Synthesis of Latent Gaussian Tree} \label{sec:achievable}

In what follows we provide the achievable rate regions to synthesize
the Gaussian tree statistics $p_{\mathbf{XY}}$ for three  distinct
cases that together cover all possible varieties that may happen in
latent Gaussian tree structures.
As we see, such intuitive classification of Gaussian trees results in better understanding the synthesis scheme for each category.
\subsection{A Basic Case Study}

In this case, we assume that the nodes at each layer are only connected to the nodes at upper/lower layers.
In other words, there is no edge between the nodes at the same layer, and they are connected to each other through one or several nodes at the upper layers.
Moreover, by deleting all the nodes at the lower layer, all the nodes at current layer should become leaves.
To better clarify our approach, it is best to begin the synthesis discussion by several illustrative examples.
\begin{ex} \label{ex:star}
Consider a latent star topology with Gaussian source $Y^{(1)}$ and sign input $B^{(1)}$, with corresponding output vector $\mathbf{X}=[X_1,X_2,...,X_n]$. This can be modeled as
\begin{align} \label{eq:regression_star}
\begin{bmatrix}
X_{1,t}\\
X_{2,t}\\
\vdots\\
X_{n,t}
\end{bmatrix}
=
\begin{bmatrix}
\alpha_1\\
\alpha_2\\
\vdots\\
\alpha_n
\end{bmatrix}
B^{(1)}_t
Y^{(1)}_t
+
\begin{bmatrix}
Z_{1,t}\\
Z_{2,t}\\
\vdots\\
Z_{n,t}
\end{bmatrix}
,~t\in\{1,2,...,N\}
\end{align}

A special case for such broadcast channel is shown in Figure \ref{fig:broadcast}, where the channel has only three outputs $X_1$, $X_2$, and $X_3$.
In the following Corollary we provide the achievable rate region for the broadcast channel
As we will show later, this is a special case in Theorem \ref{thm:achievability_basic}, which is due to soft covering lemma and the results in \cite{cuff}.
\begin{cor} \label{cor:achievability_star}
{\it
For the latent star topology characterized by \eqref{eq:regression_star}, the following rates are achievable,
\begin{align} \label{eq:conditions}
R_{Y^{(1)}}+R_{B^{(1)}}\geq I(\mathbf{X};Y^{(1)},B^{(1)})\notag\\
R_{Y^{(1)}}\geq I(\mathbf{X};Y^{(1)})
\end{align}
}
\end{cor}
Note that the sum of the rates $R_{Y^{(1)}}+R_{B^{(1)}}$ is lower bounded by $I(\mathbf{X};Y^{(1)},B^{(1)})$, which by Theorem \ref{thm:fixed} is fixed.
However, the minimum rate for $R_{Y^{(1)}}$ is achieved by $min_{p_{Y^{(1)}}} I(\mathbf{X};Y^{(1)})$.
Also due to Theorem \ref{thm:uniform_sign} we know that the optimal solution occurs when $B^{(1)}$ is uniformly distributed, i.e., $\pi_1=1/2$.

In the synthesis scheme we first need to generate the proper codebook that satisfies the rate conditions in Corollary \ref{cor:achievability_star}.
We generate $2^{NR_{B^{(1)}}}$ codewords to form the codebook $C_{B^{(1)}}$ with proper size for sign variables.
Similarly, we generate $2^{NR_{Y^{(1)}}}$ Gaussian codewords to form the codebook $C_{B^{(1)}}$.
Note that in general the latent variables have mixture Gaussian distributions, hence, such star tree is a very special case with only one \textit{Gaussian} latent variable.
Now, to obtain a Gaussian output sequence, each time we randomly pick codewords $(y^{(1)})^N$ and $(b^{(1)})^N$ from $C_{Y^{(1)}}$ and $C_{B^{(1)}}$, respectively.
Based on the observed sign instances $b^{(1)}_t,~t\in [1,...,N]$ at each time slot, we decide which channel $P_{\mathbf{X}_t|y^{(1)}_t b^{(1)}_t}$ is used to send each $y^{(1)}_t$ to generate the output $X_i^N$.
Figure \ref{fig:encoding_star} shows the synthesis scheme for this case.
\begin{figure} [h!]
\centering 
\includegraphics[width=0.8\columnwidth]{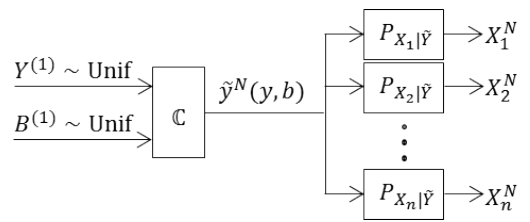}
\caption{The synthesis scheme for a latent Gaussian star tree\label{fig:encoding_star}} 
\end{figure}

We may compute the synthesized output as follows,
\begin{align}
&q(\mathbf{x}_1,...,\mathbf{x}_N)=\notag\\
&\dfrac{1}{M_{\mathbf{B}^{(1)}}}\dfrac{1}{M_{\mathbf{Y}^{(1)}}}\sum_{i=1}^{M_{\mathbf{Y}^{(1)}}}\sum_{k=1}^{M_{\mathbf{B}^{(1)}}}\prod_{t=1}^N p_{\mathbf{X}|\mathbf{Y},\mathbf{B}}(\mathbf{x}_t|\mathbf{y}^{(1)}_t[i]\mathbf{b}^{(1)}_t[k])
\end{align}
\noindent where the distribution $p_{\mathbf{X}|\mathbf{Y},\mathbf{B}}(\mathbf{x}_t|\mathbf{y}^{(1)}_t[i]\mathbf{b}^{(1)}_t[k])$ represents each channel use $t$ for corresponding input messages, and can be computed via signal model in \eqref{eq:multi-layer_linear_regression}. 

To synthesize the overall joint distribution, we need to consider the corresponding input codewords $(y^{(1)})^N$ and $(b^{(1)})^N$.
In particular, each synthesized output vector $\mathbf{X}^N$ has its own associated input codeword, and we need to pair both of these codewords to achieve the synthesized statistics.
Note that such \textit{pairing} strategy is essential, since otherwise the synthesized statistics will not be arbitrarily close to the desired distribution.
This will be further discussed in the following examples, where the correspondence between the codewords at each layer should be maintained.
\end{ex}
\begin{ex} \label{ex:dumbbell}

Consider the channel shown in Figure \ref{fig:B}(a). In this case, we are given two hidden inputs $Y_1^{(1)}$ and $Y_2^{(1)}$, and by previous arguments we know $\mathbf{B}^{(1)}=\{B_1^{(1)},B_2^{(1)},B_{12}\}$ with $B_{12}=B_1^{(1)}B_2^{(1)}$, completely determined by independent sign variables $B_1^{(1)}$ and $B_2^{(1)}$.
We may write,
\begin{align}
\begin{bmatrix}
X_{1,t}\\
X_{2,t}\\
X_{3,t}\\
X_{4,t}
\end{bmatrix}
=
\begin{bmatrix}
\alpha_{11}B^{(1)}_{1,t} & 0\\
\alpha_{21}B^{(1)}_{1,t} & 0\\
0 & \alpha_{32}B^{(1)}_{2,t}\\
0 & \alpha_{42}B^{(1)}_{2,t}
\end{bmatrix}
\begin{bmatrix}
Y^{(1)}_{1,t}\\
Y^{(1)}_{2,t}
\end{bmatrix}
+
\begin{bmatrix}
Z_{1,t}\\
Z_{2,t}\\
Z_{3,t}\\
Z_{4,t}
\end{bmatrix}
\end{align}
\noindent where $t\in\{1,2,...,N\}$ denotes each channel use.
Here, two inputs $Y_1^{(1)}$ and $Y_2^{(1)}$ are dependent and their pairwise correlation can be computed via $E[Y_1^{(1)}Y_2^{(1)}]=\gamma_{12}B_{12}=\gamma_{12}B_1^{(1)}B_2^{(1)}$, in which $\gamma_{12}$ determines the degree of correlation and is learned by certain inference algorithms, e.g., RG or CLRG \cite{mit}.
Note that the dependency relation of symbols $Y_{1,t}^{(1)}$ and $Y_{2,t}^{(1)}$ follows a Gaussian mixture model, since their covariance is a function of binary inputs $B_{1,t}^{(1)}$ and $B_{2,t}^{(1)}$.
But, note that in a given codebook consisting of $M_{\mathbf{Y}^{(1)}}$ codewords, for each realization of $\mathbf{b}_{1,t}^{(1)}\mathbf{b}_{2,t}^{(1)}$ the joint density of $\mathbf{Y}_{t}^{(1)}$ is Gaussian. 
Hence, one may divide the codebook $\mathbb{C}$ into two parts $\mathbb{S}_i,~i\in\{1,2\}$, in which each part follows a specific Gaussian density with covariance values $E[Y_{1,t}^{(1)}Y_{2,t}^{(1)}]=\gamma_{12}b_{1,t}^{(1)}b_{2,t}^{(1)}$.
In particular, to generate a codeword we first generate the codebooks $C_{\mathbf{B}^{(1)}}$ and $C_{\mathbf{Y}^{(1)}}$.
Note that the generated codewords $(\mathbf{Y}^{(1)})^N\in C_{\mathbf{Y}^{(1)}}$ are mixture Gaussians even at each time slot $t$.
To be precise, at each time slot $t$, we generate two random Gaussian sample vectors, one with $E[Y_{1,t}^{(1)}Y_{2,t}^{(1)}]=\gamma_{12}$ and the other with $E[Y_{1,t}^{(1)}Y_{2,t}^{(1)}]=-\gamma_{12}$.
Then, similar to the previous example, at synthesis step and based on the picked sign codeword, we decide which of the two sample vectors should be chosen.
The achievable region can be obtained from \eqref{eq:conditions}, and by replacing $Y^{(1)}$ with $\{Y^{(1)}_1,Y^{(1)}_2\}$ and $B^{(1)}$ with $\{B^{(1)}_1,B^{(1)}_2\}$. 
Similarly, by Theorem \ref{thm:uniform_sign} we may conclude that the optimal solution $(\pi^*_1,\pi^*_2)$ to $arg\min_{\pi_1,\pi_2} I(\mathbf{X};\mathbf{Y})$ is at $(1/2,1/2)$.
\end{ex}

\begin{figure} [h!]
\centering 
\includegraphics[width=0.9\columnwidth]{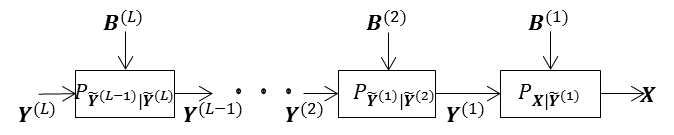}
\caption{Multi-layered output synthesis\label{fig:layered}} 
\end{figure}

Let us address more general cases, where we are having a multi-layered latent Gaussian tree with no edge between the variables at the same layer.
In other words, the variables at each layer are conditionally independent of each other given the variables at their upper layer. 
Moreover, by deleting all the nodes at the lower layer, all the nodes at current layer should become leaves.
This, in turn forms a \textit{hyper-chain} structure for latent Gaussian tree, where the \textit{hyper-nodes} consist of every variable at the same layer, and \textit{hyper-edges} are the collection of links connecting each the nodes at each adjacent layer.
Figure \ref{fig:layered} shows the general synthesis scheme.
At each layer $i$, we define $\tilde{\mathbf{Y}}^{(i)}=\{\mathbf{Y}^{(i)},\mathbf{B}^{(i)}\}$ to be the combination of input vectors.
This situation is a little more subtle than the previous single-layered cases, since we need to be more cautious on specifying the rate regions as well as the synthesis scheme.

\begin{ex} \label{ex:multi_layer}
To clarify, consider the case shown in Figure \ref{fig:B}(b), in which the Gaussian tree has two layers of inputs. 
Similar as previous cases we can compute the pairwise covariance between inputs at the first layer as $E[Y_{k,t}^{(1)}Y_{l,t}^{(1)}]=\gamma_{kl}B_{k,t}^{(1)}B_{l,t}^{(1)}$, in which $k\neq l\in\{1,2,3,4\}$. 
By the previous example, we know that the input vector $\mathbf{Y}_{t}^{(1)}$ is Gaussian for each realization of $\mathbf{B}_{t}^{(1)}=\{\mathbf{b}_{1,t}^{(1)},\mathbf{b}_{2,t}^{(1)},\mathbf{b}_{3,t}^{(1)},\mathbf{b}_{4,t}^{(1)}\}$. 
Hence, one may divide the codebook $\mathbb{C}$ into $2^{k_1}=16$ parts $\mathbb{S}_i,~i\in\{1,2,...,16\}$, in which each part follows a specific Gaussian density with covariance values $E[Y_{k,t}^{(1)}Y_{l,t}^{(1)}]=\gamma_{kl}b_{k,t}^{(1)}b_{l,t}^{(1)},~k\neq l\in\{1,2,3,4\}$.
Now, for each subset, at the second layer we are dealing with the case shown in Figure \ref{fig:B}(a), which has been resolved.
Thus, the lower bound on the achievable rates in the second layer are as follows, 
\begin{align} \label{eq:rates_fig_4b}
&R_{\mathbf{Y}^{(2)}}\geq I(\mathbf{Y}^{(1)};\mathbf{Y}^{(2)}|\mathbf{B}^{(1)})\notag\\
&R_{\mathbf{Y}^{(2)}}+R_{\mathbf{B}^{(2)}}\geq I(\mathbf{Y}^{(1)};\mathbf{Y}^{(2)},\mathbf{B}^{(2)}|\mathbf{B}^{(1)})
\end{align}
This is due to the fact that we compute subsets of codebook for each realization of $\mathbf{B}^{(1)}$.
Let us elaborate the successive codebook generation scheme in this case.

First, we need to generate the codebooks at each layer, beginning from the top layer all the way to the first layer.
The sign codebooks $C_{\mathbf{B}^{(2)}}$ and $C_{\mathbf{B}^{(1)}}$ are generated beforehand, and simply regarding the Bernoulli distributed sign vectors $\mathbf{B}^{(2)}$, and $\mathbf{B}^{(1)}$.
Hence, each sign codeword is a sequence of vectors consisting elements chosen from $\{-1,1\}$.
We may also generate the top layer codebook $C_{\mathbf{Y}^{(2)}}$ using mixture Gaussian codewords, where each codeword at each time slot consists of all possible sign realizations of $\mathbf{B}_t^{(2)}$.
Each of these settings characterize a particular Gaussian distribution for the top layer latent variables $\mathbf{Y}^{(2)}$.
The necessary number of codewords needed is $M_{\mathbf{Y}^{(2)}}=2^{NR_{\mathbf{Y}^{(2)}}}$, where the rate in the exponent is lower bounded and characterized using \eqref{eq:rates_fig_4b}.
To form the second codebook $C_{\mathbf{Y}^{(1)}}$, we know that we should use the codewords in $C_{\mathbf{Y}^{(2)}}$.
We randomly pick codewords from $C_{\mathbf{Y}^{(2)}}$ and $C_{\mathbf{B}^{(2)}}$.
Now, based on the chosen sign codeword, we form the sequence $(\mathbf{y}^{(2)}|\mathbf{b}^{(2)})^N$ to be sent through the channels.
The sign vector $\mathbf{B}^{(1)}$ consists of $k_1=4$ sign variables, hence, resulting in $2^{k_1-1}=8$ different channel realizations.
Hence, we pass the chosen sequence through $8$ different channels $p_{\mathbf{Y}^{(1)}|\tilde{\mathbf{Y}}^{(2)}\mathbf{b}^{(1)}}$.
This way, we send the chosen codeword through the $8$ noisy channels to produce a particular codeword in $C_{\mathbf{Y}^{(1)}}$.
It is important to note that we showed in subsection \ref{sec:multi_fixed} that although each of these channel correspond to different sign realizations of $\mathbf{B}^{(1)}$ vector, however, due to underlying latent Gaussian tree assumption they maintain the same rate.
Note that, such produced codeword is in fact a collection of Gaussian vectors, each corresponding to a particular sign realization $\mathbf{b}^{(l)}\in\mathbf{B}^{(1)}$.
We iterate this procedure $M_{\mathbf{Y}^{(1)}}$ times to produce enough codewords that are needed for synthesis requirements of the next layer.
The necessary size of $M_{\mathbf{Y}^{(1)}}$ is lower bounded by Corollary \ref{cor:achievability_star}.

Figure \ref{fig:encoding_general}, shows the described synthesis procedure.
In order to produce an output sequence, all we need to do is to randomly pick codewords from $C_{\mathbf{Y}^{(1)}}$ and $C_{\mathbf{B}^{(1)}}$.
Then, depending on each time slot sign realization $\mathbf{b}_t^{(1)}$ we use the corresponding channel $p_{\mathbf{X}|\mathbf{Y}^{(1)}\mathbf{b}_t^{(1)}}$ to generate a particular output sequence $\mathbf{X}^N$.
\begin{figure} [h!]
\centering 
\includegraphics[width=0.9\columnwidth]{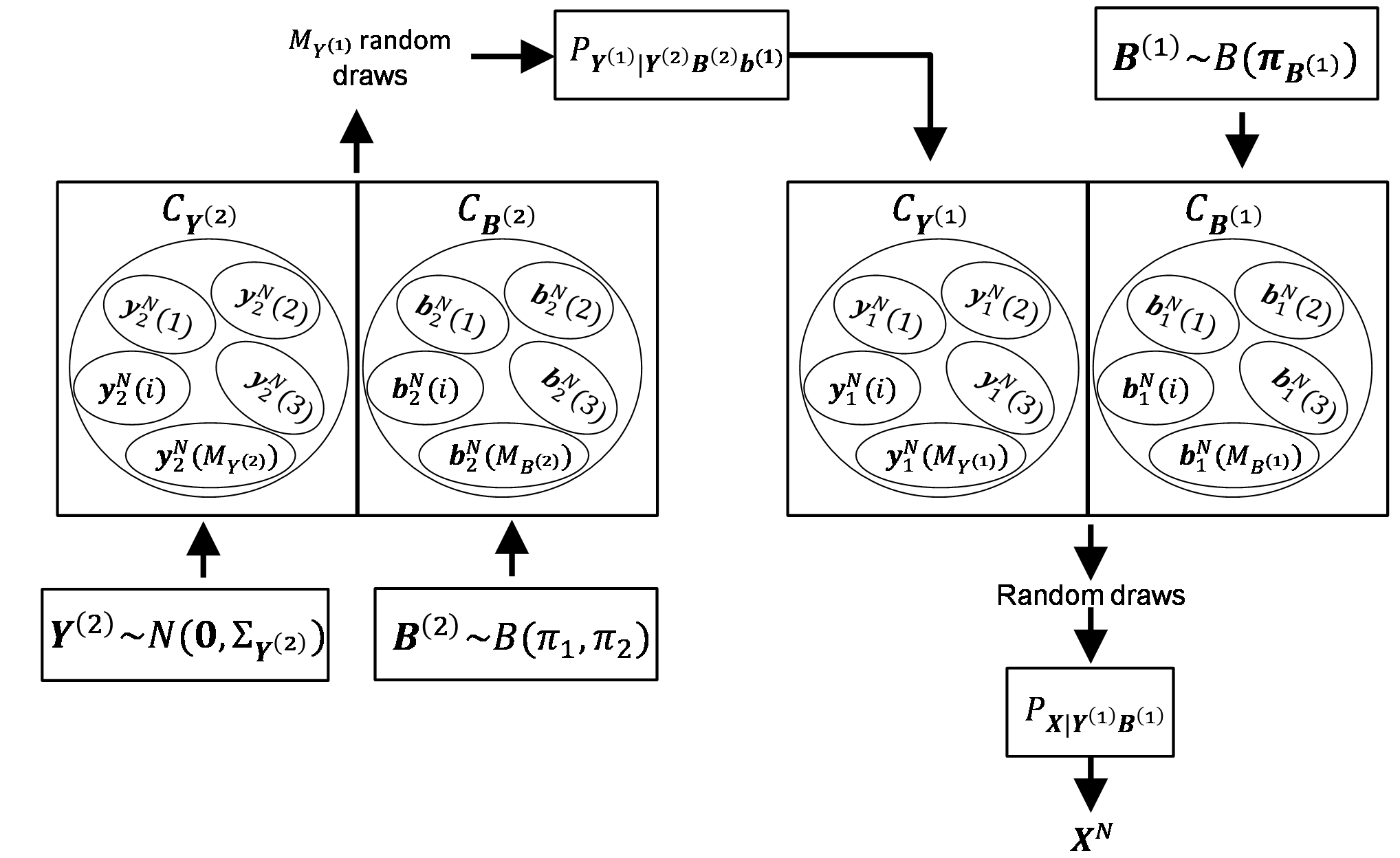}
\caption{The proposed codebook generation scheme used for a Gaussian tree shown in \ref{fig:B}(b). 
The codebook size $|C_{\mathbf{Y}^{(2)}}|$ at the top layer with shown corresponding distribution is determined by the input-output mutual information in the channel $p_{\mathbf{Y}^{(1)}|\mathbf{Y}^{(2)}\mathbf{B}^{(2)},\mathbf{b}^{(1)}}$. 
To obtain $C_{\mathbf{Y}^{(1)}}$, we randomly pick codewords from $C_{\mathbf{Y}^{(2)}}$ and $C_{\mathbf{B}^{(2)}}$ to construct $(\mathbf{Y}^{(2)}|\mathbf{B}^{(2)})^N$. Then, we send it through the eight channels $p_{\mathbf{Y}^{(1)}|\mathbf{Y}^{(2)}\mathbf{B}^{(2)},\mathbf{b}^{(1)}}$ to obtain a particular codeword $(\mathbf{Y}^{(1)})^N$.
\label{fig:encoding_general}
} 
\end{figure}

In general, the output at the $l$-th layer $\mathbf{Y}^{(l)}$ is synthesized by $\mathbf{Y}^{(l+1)}$ and $\mathbf{B}^{(l+1)}$, which are at layer $l+1$, and through different channel realizations through $\mathbf{B}^{(l)}$.
Algorithm \ref{alg:codebook} shows the general codebook generation procedure for a Gaussian tree with $L$ layers.

\begin{figure}[!h] 
 \removelatexerror
 \RestyleAlgo{boxruled}
  \begin{algorithm}[H]
   \caption{Codebook Generation for each layer of latent Gaussian tree with $L$ layers \label{alg:codebook}}
   \textbf{Input:} The needed codebook sizes $M_{\mathbf{Y}^{(l)}}$ and $M_{\mathbf{B}^{(l)}}$ for $l\in[1,L]$\\
   \textbf{Output:} The generated codebooks for each layer $l$\\
   \For{$l :=L$ to $1$}
    {
      \For{$i :=1$ to $M_{\mathbf{B}^{(l)}}$}
      {
        Randomly generate sign codewords $(\mathbf{b}^{(l)})^N$ to form $C_{\mathbf{B}^{(l)}}$\;
      }
    }
   \For{$i :=1$ to $M_{\mathbf{Y}^{(L)}}$}
      {
	\For{$\mathbf{b}^{(L)} \in \mathbf{B}^{(L)}$}
		 {
		   Randomly generate sign codewords $(\mathbf{y}^{(L)})^N|\mathbf{b}^{(L)}$\;
		 }   
		The codewords $(\mathbf{y}^{(L)})^N = \cup_{\mathbf{b}^{(L)}} (\mathbf{y}^{(L)})^N|\mathbf{b}^{(L)}$ form $C_{\mathbf{Y}^{(L)}}$     \;
      }
   \For{$l := L-1$ to $1$}
   {
      \For{$i :=1$ to $M_{\mathbf{Y}^{(l)}}$}
      {
      	Randomly pick a codeword $(\mathbf{y}^{(l+1)})^N$ from $C_{\mathbf{Y}^{(l+1)}}$\;
      	Randomly pick a codeword $(\mathbf{b}^{(l+1)})^N$ from $C_{\mathbf{B}^{(l+1)}}$\;
      	Form the combined codeword $(\mathbf{y}^{(l+1)}|\mathbf{b}^{(l+1)})^N$\;
      	\For{$\mathbf{b}^{(l)} \in \mathbf{B}^{(l)}$}
      	  {
      	    Send the codeword $(\mathbf{y}^{(l+1)}|\mathbf{b}^{(l+1)})^N$ through the channel $P_{\mathbf{Y}^{(l)}|\mathbf{Y}^{(l+1)},\mathbf{B}^{(l+1)},\mathbf{b}^{(l)}}$ to obtain $(\mathbf{y}^{(l)})^N|\mathbf{b}^{(l)}$\;
      	  }
     The codewords $(\mathbf{y}^{(l)})^N = \cup_{\mathbf{b}^{(l)}} (\mathbf{y}^{(l)})^N|\mathbf{b}^{(l)}$ form $C_{\mathbf{Y}^{(l)}}$      
      }
   }
  \end{algorithm}
\end{figure}

Therefore, to synthesize the Gaussian tree statistics that is close enough to the true Gaussian tree distribution, we first need to generate the top layer codebook $C_{\mathbf{Y}^{(l)}}$, and the sign codebooks $C_{\mathbf{B}^{(l)}},~l\in [1,L]$.
Note that the independent Gaussian noises are needed in our synthesis scheme as given source of randomness.
In Theorem \ref{thm:achievability_basic}, whose proof can be found in Appendix \ref{app:achievability_basic} we obtain the achievable rate region for multi-layered latent Gaussian tree, while taking care of sign information as well, i.e., at each layer dividing a codebook into appropriate sub-blocks capturing each realization of sign inputs.
\begin{thm} \label{thm:achievability_basic}
{\it
For a latent Gaussian tree having $L$ layers, and forming a hyper-chain structure, the achievable rate region is characterized by the following inequalities for each layer $l$,
\begin{align} \label{eq:thm_achievability_basic}
&R_{\mathbf{B}^{(l+1)}}+R_{\mathbf{Y}^{(l+1)}} \geq I[\mathbf{Y}^{(l+1)},\mathbf{B}^{(l+1)};\mathbf{Y}^{(l)}|\mathbf{B}^{(l)}]\notag\\
&R_{\mathbf{Y}^{(l+1)}}\geq I[\mathbf{Y}^{(l+1)};\mathbf{Y}^{(l)}|\mathbf{B}^{(l)}],~l\in[0,L-1]
\end{align}
}
\end{thm}
\noindent where $l=0$ shows the observable layer, in which there is no conditioning needed, since the output vector $\mathbf{X}$ is already assumed to be Gaussian.
Notice that using Theorem \ref{thm:fixed}, we can partially characterize the first lower bound on the sum of rates, since this is a fixed quantity, given the observables covariance matrix; however, analytically characterizing the lower bound on each of the rates $R_{\mathbf{Y}^{(l+1)}}$ due to presence of mixture Gaussian inputs $\mathbf{Y}^{(l+1)}$ is a hard problem to solve.
We refer the reader to \cite{pi} for further results on mixture Gaussian variables.
\end{ex}

\begin{figure} [h!]
\centering
\includegraphics[width=0.95\columnwidth]{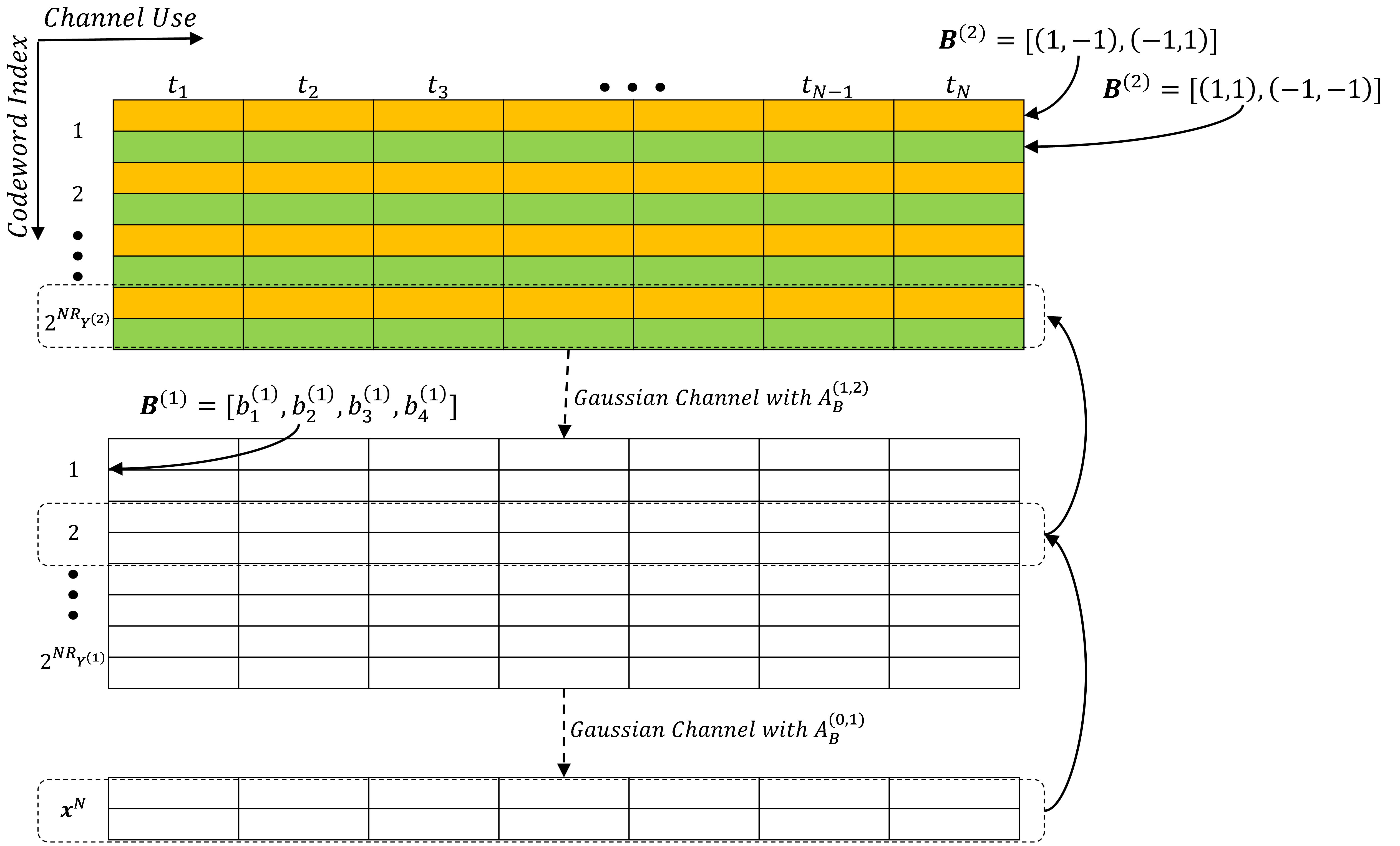} 
\caption{Synthesis approach for the Gaussian tree in Figure \ref{fig:B2}}\label{fig:synthesis_2b} 
\end{figure}

Let us assume using the top-down approach shown in Algorithm \ref{alg:codebook} we generate the appropriate codebooks at each layer.
To pick appropriate sample codeword, each time we need to keep track of input-output codewords relationship.
Considering each particular layer outputs, we keep track of the corresponding input codeword that generated such output.
For instance, consider the same Gaussian tree shown in Figure \ref{fig:synthesis_2b}.
To generate an output sequence $\mathbf{x}^N$, we randomly pick two codewords $(\mathbf{y}^{(1)})^N$ and $(\mathbf{b}^{(1)})^N$ from the corresponding codebooks. The sign codeword decides which channel to be used in order to obtain the outputs.
Hence, there is a correspondence between such input codewords and the generated outputs.
Similarly, the codeword $(\mathbf{y}^{(1)})^N$ is an output of the top layer inputs, generated by randomly chosen codewords $(\mathbf{y}^{(2)})^N$ and $(\mathbf{b}^{(2)})^N$.
Figure \ref{fig:synthesis_2b} shows the synthesis scheme that is proposed for the two-layered latent Gaussian tree shown in Figure \ref{fig:B2}.

In Figure \ref{fig:synthesis_2b}, different colors in codebooks correspond to different sign realizations. For example, as we know the top layer sign inputs $\mathbf{B}^{(2)}$ can have $2^{k_2-1}=2$ different sign realizations, hence $2$ different colors are shown in the corresponding codebook.
Also, note that each cell in codebooks may contain a vector of samples, due to the fact that each layer usually contains more than one variable.
For example, each cell in the top codebook contains two Gaussian samples, corresponding to $\mathbf{y}^{(2)}_t|\mathbf{b}^{(2)}_{t}$ where $\mathbf{y}^{(2)} = [y_1^{(2)},y_2^{(2)}]$.
The bottom-up synthesis approach first randomly picks the sequences $(\mathbf{y}^{(1)})^N$ and $(\mathbf{b}^{(1)})^N$ from $C_{\mathbf{Y}^{(1)}}$ and $C_{\mathbf{B}^{(1)}}$, respectively, and forms $(\mathbf{y}^{(1)}|\mathbf{b}^{(1)})^N$, then finds the corresponding input codeword  $(\mathbf{y}^{(2)}|\mathbf{b}^{(2)})^N$ that generated such codeword at the first layer.
Then, the chosen codeword $(\mathbf{y}^{(1)}|\mathbf{b}^{(1)})^N$ is used to generate the output vector $\mathbf{x}^N$ through the given Gaussian channel.
The sequence of samples in this case (as shown in Figure \ref{fig:synthesis_2b}) is $[\mathbf{x}^N,(\mathbf{y}^{(1)}|\mathbf{b}^{(1)})_2^N,(\mathbf{y}^{(2)}|\mathbf{b}^{(2)})_{M_{Y^{(2)}}}^N]$. Remember that each layer's codeword carries its corresponding sign information characterized in codebook generation step.

In general, this procedure should always hold from the bottom to top of latent Gaussian tree, in order to keep a valid joint dependency among the variables at every layer.
Algorithm \ref{alg:synthesis} shows this procedure for any general Gaussian tree.
\begin{figure}[!h] 
 \removelatexerror
 \RestyleAlgo{boxruled}
  \begin{algorithm}[H]
   \caption{Synthesis approach for a latent Gaussian tree with $L$ layers \label{alg:synthesis}}
   \textbf{Input:} Generated codebooks from Algorithm \ref{alg:codebook}\\
   \textbf{Output:} A valid sequence $(\mathbf{x}^N,(\mathbf{y}|\mathbf{b})^N)$ from the synthesized Gaussian distribution\\
   Randomly pick the codewords $(\mathbf{y}^{(1)})^N$ and $(\mathbf{b}^{(1)})^N$ from $C_{\mathbf{Y}^{(1)}}$ and $C_{\mathbf{B}^{(1)}}$, respectively, and form $(\mathbf{y}^{(1)}|\mathbf{b}^{(1)})^N$\;
   \For{$l := 1$ to $L-1$}
   {
      For each codeword $(\mathbf{y}^{(l)}|\mathbf{b}^{(l)})^N$, pick the corresponding codeword $(\mathbf{y}^{(l+1)}|\mathbf{b}^{(l+1)})^N$, which has been employed generating it \;
    }
    For $(\mathbf{y}^{(L-1)}|\mathbf{b}^{(L-1)})^N$, pick the corresponding codeword $(\mathbf{y}^{(L)}|\mathbf{b}^{(L)})^N$ at layer $L$\\
    Send the chosen codeword $(\mathbf{y}^{(1)}|\mathbf{b}^{(1)})^N$ through the channel $p_{\mathbf{X}|\mathbf{Y}^{(1)}\mathbf{B}^{(1)}}$ to obtain $\mathbf{x}^N$\\
   Output the overall sequence of codewords $[\mathbf{x}^N,(\mathbf{y}^{(1)}|\mathbf{b}^{(1)})^N,...,(\mathbf{y}^{(L)}|\mathbf{b}^{(L)})^N]$ \;
  \end{algorithm}
\end{figure}

\subsection{The case with observables adjacent with more than one latent variable} \label{sub_sec:internal}
Here, we consider more general cases, which may allow nodes at each layer to have more than one neighbor from upper layer. This way, by deleting the nodes at the lower layer, we may end up with several internal (non-leaf) nodes at the current layer.
We need to propose a revised achievability proof to characterize the achievable rate region.
To clarify our approach, consider the following example shown in Figure \ref{fig:internal}.
\begin{ex}
This is a double layer latent Gaussian tree, with $X_6$ as an internal node. As it can be seen, after the first step by summing out $X_6$, we created a clique at the next layer. This is not a latent Gaussian tree structure anymore. In fact, this can be seen as a \textit{junction tree} structure. The problem with such structure is that, given the nodes at upper layer, i.e, $Y_1^{(2)}$, the nodes at the lower layer, i.e, $Y^{(1)}_i,~i\in [2,4]$ are not conditionally independent anymore. 
This, violates some of the constraints in the acheivability results in Theorem \ref{thm:achievability_basic}.
\begin{figure} [h!]
\centering
\includegraphics[width=0.8\columnwidth]{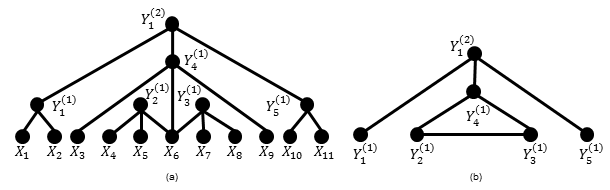} 
\caption{$(a)$ The original two layered Gaussian tree $(b)$ Obtained grpah after the first iteration}\label{fig:internal} 
\end{figure}

To address such problem, we introduce another \textit{latent pseudo-node}, $Y^{(2)}_2$ and connect it to all the nodes forming the clique, as it is shown in Figure \ref{fig:middle_step}.
\begin{figure} [h!]
\centering
\includegraphics[width=0.3\columnwidth]{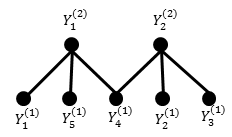} 
\caption{The intermediate step needed to address the internal node issue: By adding a pseudo node $Y^{(2)}_2$ we break the clique and trun it into a tree structure again}\label{fig:middle_step} 
\end{figure}

Note that we certainly can represent the formed clique via a latent Gaussian tree (star tree), since from the first iteration we already know that the hidden nodes at the first layer are connected to each other through $X_6$. So, now after adding the \textit{pseudo-node} $Y^{(2)}_2$, we know that the following equality holds for the new correlations $\rho_{y^{(2)}_2y^{(1)}_i}=\rho_{x_6y^{(1)}_i},~i\in [2,4]$. Hence, we may see $Y^{(2)}_2$ as the \textit{mirror} node to $X_6$, which is added to the nodes in the second layer.
Finally, we only need to update the set of nodes at the second layer to $Y^{(2)}=\{Y^{(2)}_1,Y^{(2)}_2\}$.
\begin{rmk}
In general, the synthesis scheme will remain the same as the basic case, with one tweak: at each layer $l+1$ we may need to perform an intermediate step, in which we transform cliques into latent trees (star structure), by adding enough pseudo-nodes to the set of upper layer latent nodes.
The sufficient number of pseudo nodes to be added should be the same as the number of internal nodes at layer $l$.
Through such procedure, due to the addition of new nodes (the pseudo nodes) both the corresponding rates $R_{\mathbf{Y}^{(l+1)}}$ and $R_{\mathbf{B}^{(l+1)}}$, and consequently the achievable rate regions will be changed to the following.
\begin{align}
&R_{\mathbf{B'}^{(l+1)}}+R_{\mathbf{Y'}^{(l+1)}} \geq I[\mathbf{Y'}^{(l+1)},\mathbf{B'}^{(l+1)};\mathbf{Y}^{(l)}|\mathbf{B}^{(l)}]\notag\\
&R_{\mathbf{Y'}^{(l+1)}}\geq I[\mathbf{Y'}^{(l+1)};\mathbf{Y}^{(l)}|\mathbf{B}^{(l)}]
\end{align}

\noindent where $\mathbf{Y'}^{(l+1)}=\mathbf{Y}^{(l+1)}\cup \mathbf{Y}^{(l)}_p$ and $\mathbf{B'}^{(l+1)}=\mathbf{B}^{(l+1)}\cup \mathbf{B}^{(l)}_p$ are the new input vectors at layer $l+1$, with $\mathbf{Y}^{(l)}_p$ and $\mathbf{B}^{(l)}_p$, being the newly added psuedo latent and sign inputs.
\end{rmk}
\end{ex}

\subsection{The case with observables at different layers}
\begin{figure} [h!]
\centering
\includegraphics[width=0.3\columnwidth]{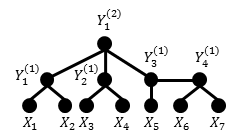} 
\caption{Latent Gaussian tree with adjacent nodes at layer $1$}\label{fig:internal2} 
\end{figure}

Consider a case where an edge is allowed between the variables at the same layer.
In this situation we violate a conditional independence constraint used in achievability proof of the basic case, since due to presence of such intra-layer links, given the upper layer inputs, the conditional independence of lower layer outputs is no longer guaranteed.
However, again by revising the proof procedure we may show the achievability results in this case as well.
To address this issue we need to reform the latent Gaussian tree structure by choosing an appropriate root such that the variables in the newly introduced layers mimic the basic scenario, i.e., having no edges between the variables at the same layer.
In particular, we begin with the top layer nodes, and as we move to lower layers we seek each layer for the adjacent nodes at the same layer, and move them to a newly added layer in between the upper and lower layers.
In this way, we introduce new layers consisting of those special nodes, but this time we are dealing with a basic case.
Note that such procedure might place the output variables at different layers, i.e., all the output variables are not generated using inputs at a single layer.
We only need to show that using such procedure and previously define achievable rates, one can still simulate output statistics with vanishing total variation distance.
To clarify, consider the following example in Figure \ref{fig:internal2}.

\begin{ex}
As it can be seen, there are two adjacent nodes in the first layer, i.e., $Y_3^{(1)}$ and $Y_4^{(1)}$ are connected.
Using the explained procedure, we may move $Y_4^{(1)}$ to another newly introduced layer, then we relabel the nodes again to capture the layer orderings.
The reformed Gaussian tree is shown in Figure \ref{fig:internal3}.
In the new ordering, the output variables $X_6$ and $X_7$ will be synthesized one step after other inputs.
The input $Y_1^{(3)}$ is used to synthesize the outputs vector $\mathbf{Y}^{(2)}$, which such vector used to generate the first layer outputs, i.e., $X_1$ to $X_5$ and $Y_1^{(1)}$.
At the last step, the input $Y_1^{(1)}$ will be used to simulate the output pair $X_6$ and $X_7$.
By Theorem \ref{thm:achievability_basic} we know that both simulated densities regarding to $q_{X^N_1,X^N_2,X^N_3,X^N_4,X^N_5}$ and $q_{X^N_6,X^N_7}$ approach to their corresponding densities as $N$ grows.
We need to show that the overall simulated density $q_{\mathbf{X}^N}$ also approaches to $\prod_{t=1}^N p_{\mathbf{X}}(\mathbf{x}_t)$ as well.

We need to be particularly cautious in keeping the joint dependency among the generated outputs at different layers: For each pair of outputs $(X^N_6,X^N_7)$, there exists an input codeword $(Y_1^{1})^N$, which corresponds to the set of generated codewords $(X^N_1,X^N_2,X^N_3,X^N_4,X^N_5)$, where together with $(Y_1^{1})^N$ they are generated using the second layer inputs.
Hence, in order to maintain the overall joint dependency of the outputs, we always need to match the correct set of outputs $X_1^N$ to $X_5^N$ to each of the output pairs $(X^N_6,X^N_7)$, where this is done via $(Y_1^{1})^N$.
\begin{figure} [h!]
\centering
\includegraphics[width=0.3\columnwidth]{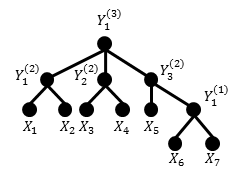} 
\caption{Another layer introduced to address the issue}\label{fig:internal3} 
\end{figure}

In general, we need to keep track of the indices of generated output vectors at each layer and match them with corresponding output vector indices at other layers.
This is shown in Lemma \ref{lem:achievability_general2}, whose proof can be found in Appendix \ref{app:achievability_general2},
\begin{lem} \label{lem:achievability_general2}
{\it
For a latent Gaussian tree having $L$ layers, and not containing an internal node at any iteration, by rearranging each layer so that there is no intra-layer edges, the achievable rate region at each layer $l$ is characterized by the same inequalities as in \eqref{eq:thm_achievability_basic}.
}
\end{lem}
\end{ex}

By Lemma \ref{lem:achievability_general2} we may easily extend our results and show that, interestingly, to generate a latent Gaussian tree, we only need its top layer nodes acting as common random sources (and independent Gaussian and Bernoulli noises) to synthesize the entire Gaussian tree structure.
Algorithms \ref{alg:codebook} and \ref{alg:synthesis} are used for the reformed structure for codebook generation and synthesis steps, respectively.
\begin{cor}
{\it
Given any latent Gaussian tree consisting of $L$ hidden layers along with an output vector $\mathbf{X}$, by combining the aforementioned procedures described in the last three subsections and using the top layer inputs, i.e., the inputs at the $L$-th layer, the independent Gaussian noises, and the independent Bernoulli variables, the entire set of nodes in a latent Gaussian tree can be synthesized if the rates at each layer satisfy the constraints captured in \eqref{eq:thm_achievability_basic}.
}
\end{cor}

Note that, the top layer nodes, without considering any other node in a tree, will certainly form a chain (or a single node in a special case) structure.

\section{Conclusion} \label{sec:conclusion}
In this paper, we formulated a synthesis problem through layered forwarding channels to synthesize those statistics that characterize the latent Gaussian tree structures.
Then we deduced an interesting conclusion under which maximizing the achievable rate region also resulted in quantifying the maximum amount of lost information on pairwise correlation signs.
Through three different cases we found the achievable rate regions to correctly synthesize the Gaussian outputs, satisfying specific set of constraints.
Our layered synthesis approach is shown to be efficient and accurate in terms of reduced required number of parameters needed to synthesize the output statistics, and its closeness to the desired statistics in terms of their total variation distance.

\bibliography{reference}
\bibliographystyle{IEEEtran}

\appendices

\section{Proof of Theorem \ref{thm:B}} \label{app:B}

\begin{figure} [h!]
\centering
\includegraphics[scale=0.5]{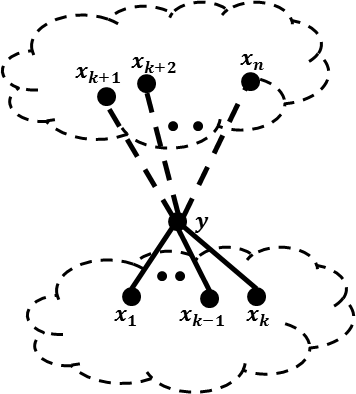} 
\caption{Neighborhood of hidden variable $y$}\label{fig:neighbor} 
\end{figure}

First, let's prove the first part.
Consider the case in Figure \ref{fig:neighbor}. The hidden node $y$, has $k$ observable neighbors $\{x_1,...,x_k\}$, while it is connected through two or more edges to other observable nodes $\{x_{k+1},...,x_n\}$. 
Given only observable covariance matrix $\Sigma_x$, we can compute the empirical pairwise covariance values, hence all $\rho_{x_ixj}$ are fixed.

Without loss of generality, suppose we flip the sign of $\rho_{x_1y}$. 
To maintain the same covariance matrix $\Sigma_x$, the sign of all $\rho_{x_jy}$, $j\in\{2,...,k\}$ should be flipped. Since, we know $\rho_{x_1x_j}=\rho_{x_1y}\rho_{x_jy}$, for all $j\in\{2,...,k\}$ is fixed.
Also, the sign of all pairwise covariance values between $y$ and $x_i$, for all $i\in\{k+1,...,n\}$ should be flipped. The same argument as the previous case can be used. However, in this case, all we know is that odd number of sign-flips for the edge-weights between each $y$ and $x_i$ should happen.
Using the above arguments, we can see that all $\rho_{x_jy}$, for $j\in\{1,...,k\}$ maintain their signs, or otherwise all of their signs should be flipped.

\begin{figure} [h!]
\centering
\includegraphics[scale=0.5]{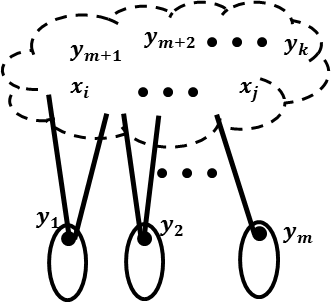} 
\caption{Figure illustrating the inductive proof}\label{fig:induction} 
\end{figure}

For the second part, 
We inductively show that given a minimal latent tree, with $n$ observable $x_1,...,x_n$ and with $k$ hidden nodes $y_1,...,y_k$, we can find $2^k$ latent trees with different edge-signs that induce the same $\Sigma_x$.
This is already shown for the star tree shown in Figure \ref{fig:broadcast}.
Suppose such claim holds for all Gaussian trees with $k'<k$ latent nodes. 
Consider an arbitrary latent tree with $k$ hidden nodes and $n$ observable.
Some of these hidden nodes certainly have leaf observable neighbors, which we group them together.
Now, note that the problem of finding equivalent sign permutations in this tree can be translated into a problem with smaller tree: Delete all of those leaf observable groups, and treat their hidden parent $y_i$ as their representative. Suppose there are $m$ hidden nodes $\{y_1,...,y_m\}$, which can represent each of these groups. This case is illustrated in Figure \ref{fig:induction}. 
Note, as depicted by this Figure, the internal observables as well as those leaf observables directly connected to them remain intact.
By replacing all of these groups with a single node $y_i$, $i\in\{1,2,...,m\}$, we obtain a smaller tree. Now, we can simply assume that all $y_1,...,y_m$ are observable and their pairwise covariance values are determined. Hence, this tree only has $k-m$ remaining hidden nodes, so due to inductive step it has $2^{k-m}$ possible equivalent trees with different edge-signs.

\begin{figure} [h!]
\centering
\includegraphics[scale=0.45]{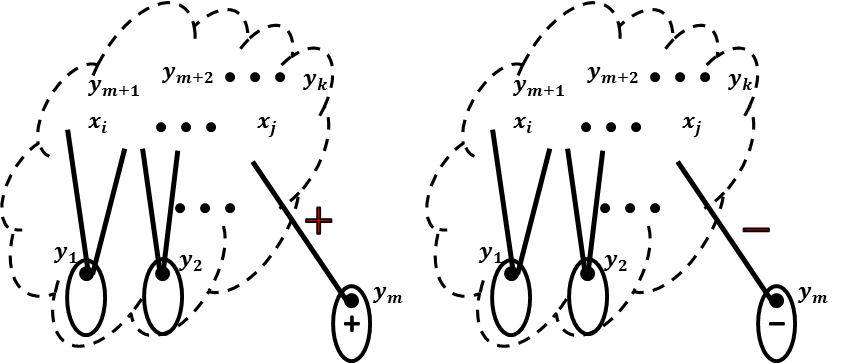} 
\caption{Obtaining $m$-th step from $m-1$-th step}\label{fig:mth} ñ
\end{figure}

It remains to show that by adding back those $m$ groups of observable, we obtain the claimed result. Add back two groups corresponding to $y_1$ and $y_2$. Now, $y_1$ and $y_2$ can be regarded as hidden nodes, so now there are $k-m+2$ hidden nodes, which due to inductive step has $2^{k-m+2}$ equivalent representations of edge-weights. This can be shown up to $m-1$-th step by adding back the groups for $y_1,...,y_{m-1}$ nodes, and having a size of $k-1$ nodes, and again due to induction having $2^{k-1}$ equivalent sign combinations.
By adding back the $m$-th group, we can obtain two equivalent classes: $b^{(m)}$ or $-b^{(m)}$, where $b^{(m)}$ shows the sign value of the $m$-th group.  
This is shown in Figure \ref{fig:mth}
Hence, we obtain $2\times 2^{k-1}=2^k$ edge-signs.

This completes the proof.

\section{Proof of Theorem \ref{thm:fixed}} \label{app:fixed}

Let's first show that the mutual information $I(\mathbf{X},\tilde{\mathbf{Y}})$ given $\Sigma_{\mathbf{x}}$ is only a function of pairwise correlations $\rho_{x_ix_j}$, for all $x_i,x_j\in\mathbf{X}$.
In a latent Gaussian tree, three cases may happen: The edges can be between two observable, an observable and a latent node, or between two latent nodes.

$(1)$ $x_i$ and $x_j$ are either adjacent or they are connected only through several observables.
In this case, since all the pairwise correlations along the path are determined given $\Sigma_{\mathbf{x}}$, so the correlation values are fixed.

$(2)$ $x_i$ and $x_j$ are not adjacent and there is at least one hidden node, e.g., $y_1$ connecting them. First, suppose $y_1$ and $x_i$ are adjacent. Since, we assume the tree is minimal, so there should be at least another observable $x_k$ that is connected (but not necessarily adjacent) to $y_1$. Hence, $y_1$ acts as a common ancestor to $x_i$, $x_j$, and $x_k$. By changing $\rho_{x_iy_1}$ to another value $\rho'_{x_iy_1}$, by equation $\rho_{x_ix_j}=\rho_{x_iy_1}\rho_{x_jy_1}$ we have to change $\rho_{x_jy_1}$ to $\rho'_{x_jy_1}=\dfrac{\rho_{x_iy_1}}{\rho'_{x_iy_1}}\rho_{x_jy_1}$. Similarly, by equality $\rho_{x_ix_k}=\rho_{x_iy_1}\rho_{x_ky_1}$, we know $\rho'_{x_ky_1}=\dfrac{\rho_{x_iy_1}}{\rho'_{x_iy_1}}\rho_{x_ky_1}$. However, by another equality $\rho_{x_jx_k}=\rho_{x_jy_1}\rho_{x_ky_1}$, we deduce $\rho'_{x_ky_1}=\dfrac{\rho_{x_jy_1}}{\rho'_{x_jy_1}}\rho_{x_ky_1}$. The obtained correlation $\rho'_{x_ky_1}$ should have the same value in both equations, hence, we deduce the equality $\dfrac{\rho_{x_iy_1}}{\rho'_{x_iy_1}}=\dfrac{\rho_{x_jy_1}}{\rho'_{x_jy_1}}$.
On the other hand, from $\rho_{x_ix_j}=\rho_{x_iy_1}\rho_{x_jy_1}$, we have $\dfrac{\rho_{x_iy_1}}{\rho'_{x_iy_1}}=\dfrac{\rho'_{x_jy_1}}{\rho_{x_jy_1}}$.
By these two equations we may conclude $\rho_{x_iy_1}=\rho'_{x_iy_1}$, a contradiction.
Hence, in this case, given $\Sigma_\mathbf{x}$ we cannot further vary the edge-weights.
Second, consider the case, where $x_i$ is connected to $y_1$ through several observables.
Then, instead of $x_i$, we can simply consider the observable that is adjacent to $y_1$, say, $x'_i$ and follows the previous steps to obtain the result.
Hence, in general if three nodes are connected to each other through separate paths and have a common ancestor $y_1$, then the pairwise correlations between the hidden nodes and each of the observables remain fixed.

$(3)$  Consider two adjacent latent nodes $y_1$ and $y_2$. By minimality assumption and having a tree structure, it can be seen that there are at least two observable for each of the latent nodes that share a common latent parent. Let's assign $x_i$ and $x_j$ to a common ancestor $y_1$ while $x_k$ and $x_k$ are descendant to $y_2$. Considering $x_i$, $x_j$, and $x_k$, who share a common parent $y_1$ ($x_k$ is connected to $y_1$ through $y_2$), using arguments on case $(2)$, we conclude that $\rho_{x_iy_1}$ and $\rho_{x_jy_1}$ should be fixed.
Similarly, we can consider $x_i$, $x_k$, and $x_l$ to show that $\rho_{x_ky_1}$ and $\rho_{x_ly_1}$ are fixed.
Now, by considering any observable pair that go through both $y_1$ and $y_2$ the result follows. For example, considering $\rho_{x_ix_k}=\rho_{x_iy_1}\rho_{y_1y_2}\rho_{x_ky_1}$, we can see that since given $\rho_{x_ix_k}$, both $\rho_{x_iy_1}$ and $\rho_{x_ky_1}$ are determined, so $\rho_{y_1y_2}$ should be determined as well.
This completes the first part of the proof.

Second, note that one may easily show that $I(\mathbf{X},\tilde{\mathbf{Y}})=1/2\log\dfrac{|\Sigma_\mathbf{x}||\Sigma_{\tilde{\mathbf{y}}}|}{|\Sigma_{\mathbf{x}\tilde{\mathbf{y}}}|}$. Now, since $p_{\mathbf{X,\tilde{Y}}}$ induces a latent Gaussian tree and $p_{\tilde{\mathbf{Y}}}$ is its marginalized density after summing out the random vector $\mathbf{X}$. By \cite{arxiv}, we know that $|\Sigma_{\mathbf{X},\tilde{\mathbf{Y}}}|=\prod_{(i,j)\in E} (1-\rho^2_{i,j})$, where $\rho_{i,j}$ are the pairwise correlations, between two adjacent variables (hidden or observable) in a latent Gaussian tree.
Now, since the observables are only leaves, by summing them out we end with another Gaussian tree consisting of only latent variables.
Thus, again by \cite{arxiv} we know $|\Sigma_{\tilde{\mathbf{Y}}}|=\prod_{(i,j)\in E_y} (1-\rho^2_{i,j})$, where $E'$ is the set of edges in the new Gaussian tree.
Observe that all the common terms of the form $(1-\rho^2_{y_iy_j})$, for some $(y_i,y_j)\in E$ will be canceled out with the terms in $|\Sigma_{\tilde{\mathbf{Y}}}|$.
Hence, the mutual information has the following form $I(\mathbf{X},\tilde{\mathbf{Y}})=1/2\log\dfrac{|\Sigma|_\mathbf{X}}{\prod_{(x_i,y_j)\in E (1-\rho^2_{x_iy_j})}}$.
Now, to find each correlation value $\rho_{x_iy_j}$, for some $X_i$ and $Y_j$,
first consider the star model, with one hidden node, and three leaves, e.g., Figure \ref{fig:broadcast}. We can write: $\rho^2_{x_1y}=\dfrac{\rho_{x_1x_2}\rho_{x_1x_3}}{\rho_{x_2x_3}}$, $\rho^2_{x_2y}=\dfrac{\rho_{x_1x_2}\rho_{x_2x_3}}{\rho_{x_1x_3}}$, and $\rho^2_{x_3y}=\dfrac{\rho_{x_1x_3}\rho_{x_2x_3}}{\rho_{x_1x_2}}$. 
For a general structure, if we replace $1\leftarrow i$, $2\leftarrow j_i$, and $3\leftarrow k_i$, we conclude that $\rho^2_{x_iy_j}=\dfrac{\rho_{x_ix_{j_i}}\rho_{x_ix_{k_i}}}{\rho_{x_{j_i}x_{k_i}}}$, for any three distinct $i,~j_i$ and $k_i$. As it may seem, there are many equations for computing $\rho^2_{x_iy_j}$, which all of these expressions should be equal, i.e., the covariance matrix $\Sigma_x$ should be representable by a given latent tree model.

\section{Proof of Theorem \ref{thm:uniform_sign}} \label{app:uniform_sign}

Suppose the latent Gaussian tree has $k$ latent variables,i.e., $\mathbf{Y}=[Y_1,Y_2,...,Y_k]$.
By adding back the sign variables the joint density $p_{\mathbf{XY}}$ becomes a Gaussian mixture model.
One may model such mixture as the summation of densities that are conditionally Gaussian, given sign vector.
\begin{align} \label{eq:f_i}
p_{\mathbf{XY}}(\mathbf{x,y}) = \sum_{i=0}^{2^k-1} \eta_{\mathbf{B}_i}f_i(\mathbf{x,y})
\end{align}

\noindent where each $\eta_{\mathbf{B}_i}$ captures the overall probability of the binary vector $\mathbf{B}_i=[b_{1i},b_{2i},...,b_{ki}]$, with $b_{ji}\in\{0,1\}$.
Here, $b_{ji}=0$ is equivalent to having $b_{ji}=-1$.
The terms $f_i(\mathbf{x,y})$ are conditional densities of the form $p(\mathbf{x,y}|\mathbf{B}_i)$

In order to characterize $I(\mathbf{X},\mathbf{Y})$, we need to find $p_{\mathbf{Y}}(\mathbf{y})$ in terms of $\eta_{\mathbf{B}_i}$ and conditional Gaussian densities as well.
First, let's show that for any two hidden nodes $y_i$ and $y_j$ in a latent Gaussian tree, we have $E[y_iy_j]=\rho_{y_iy_j}b_ib_j$. The proof goes by induction: 
We may consider the structure shown in Figure \ref{fig:B}(a) as a base, where we proved that $B_{12}=B^{(1)}_1B^{(1)}_2$.
Then, assuming such result holds for any Gaussian tree with $k-1$ hidden nodes, we prove it also holds for any Gaussian tree with $k$ hidden nodes. Let's name the newly added hidden node as $y_k$ that is connected to several hidden and/or observable such that the total structure forms a tree. 
Now, for each newly added edge we assign $b_kb_{n_k}$, where $n_k\in N_k$ is one of the neighbors of $y_k$. Note that this assignment maintains the pairwise sign values between all previous nodes,
since to find their pairwise correlations we go through $y_k$ at most once, where upon entering/exiting $y_k$ we multiply the correlation value by $b_k$, hence producing $b_k.b_k=1$, so overall the pairwise correlation sign does not change. 
Note that the other pairwise correlation signs that do not pass through $C_k$ remain unaltered.
One may easily check that by assigning $b_kb_{n_k}$ to the sign value of each newly added edge we make $y_k$ to follow the general rule, as well.
Hence, overall we showed that $E[y_iy_j]=\rho_{y_iy_j}b_ib_j$ for any $y_i,y_j\in\mathbf{Y}$.
This way we may write $\Sigma_{\mathbf{y}}=B\Sigma'_{\mathbf{y}}B$, where $\rho_{y_iy_j}\in\Sigma'_{\mathbf{y}}$ and $b_i\in B$ is $k\times k$ diagonal matrix.
One may easily see that both $B$ and its negation matrix $-B$ induce the same covariance matrix $\Sigma_\mathbf{y}$. As a result,  if we define $\eta_{\bar{\mathbf{B}}_i}$ as a compliment of $\eta_{\mathbf{B}_i}$, we can write the mixture density $p_{\mathbf{Y}}(\mathbf{y})$ as follows,
\begin{align} \label{eq:g_i}
p_{\mathbf{Y}}(\mathbf{y}) = \sum_{i=0}^{2^{k-1}-1} (\eta_{\mathbf{B}_i}+\eta_{\bar{\mathbf{B}}_i})g_i(\mathbf{y})
\end{align}

\noindent where the conditional densities can be characterized as $g_i(\mathbf{y})=p(\mathbf{y}|\mathbf{B}_i)=p(\mathbf{y}|\bar{\mathbf{B}}_i)$.
We know that $g_i(\mathbf{y})=\int f_j(\mathbf{x,y}) d\mathbf{x}$, where $j$ may correspond to either $\mathbf{B}_i$ or $\bar{\mathbf{B}}_i$.

First, we need to show that the mutual information $I(\mathbf{X},\mathbf{Y})$ is a convex function of $\eta_{\mathbf{B}_i}$ for all $i\in[0,2^k-1]$. By equality $I(\mathbf{X},\mathbf{Y})=h(\mathbf{X})-h(\mathbf{X}|\mathbf{Y})$, and knowing that given $\Sigma_{\mathbf{x}}$ the entropy $h(\mathbf{X})=1/2\log (2\pi e)^n|\Sigma_{\mathbf{x}}|$ is fixed, we only need to show that the conditional entropy $h(\mathbf{X}|\mathbf{Y})$ is a concave function of $\eta_{\mathbf{B}_i}$.
Using definition of entropy and by replacing for $p_{\mathbf{XY}}$ and $p_{\mathbf{Y}}$ using equations \eqref{eq:f_i} and \eqref{eq:g_i}, respectively, we may characterize the conditional entropy.
By taking second order derivative, we deduce the following,
\begin{align} \label{eq:second_derivative}
\dfrac{\partial^2 h(\mathbf{X}|\mathbf{Y})}{\partial^2\eta_i\eta_j} =& -\int\int\dfrac{f_i(\mathbf{x,y})f_j(\mathbf{x,y})}{p_{\mathbf{XY}}}d\mathbf{x}d\mathbf{y} \notag\\&+ \int\dfrac{\tilde{g}_i(\mathbf{y})\tilde{g}_j(\mathbf{y})}{p_{\mathbf{Y}}}d\mathbf{y}
\end{align}

\noindent where for simplicity of notations we write $\eta_i$ instead of $\eta_{\mathbf{B}_i}$. Also, $\tilde{g}_i(\mathbf{y})=\tilde{g}_{\bar{i}}(\mathbf{y})=g_i(\mathbf{y})$ for $i\in[0,2^{k-1}-1]$.
Note the following relation,
\begin{align} \label{eq:equality1}
\int\int\dfrac{\tilde{g}_i(\mathbf{y})f_j(\mathbf{x,y})p_{\mathbf{X|Y}}}{p_{\mathbf{XY}}}d\mathbf{x}d\mathbf{y} &= \int\int\dfrac{\tilde{g}_i(\mathbf{y})f_j(\mathbf{x,y})}{p_{\mathbf{Y}}}d\mathbf{x}d\mathbf{y}\notag\\
&=\int\dfrac{\tilde{g}_i(\mathbf{y})}{p_{\mathbf{Y}}}(f_j(\mathbf{x,y})d\mathbf{x})d\mathbf{y}\notag\\
&=\int\dfrac{\tilde{g}_i(\mathbf{y})\tilde{g}_j(\mathbf{y})}{p_{\mathbf{Y}}}d\mathbf{y}
\end{align}

The same procedure can be used to show, 
\begin{align} \label{eq:equality2}
\int\int\dfrac{\tilde{g}_j(\mathbf{y})f_i(\mathbf{x,y})p_{\mathbf{X|Y}}}{p_{\mathbf{XY}}}d\mathbf{x}d\mathbf{y}=\int\dfrac{\tilde{g}_i(\mathbf{y})\tilde{g}_j(\mathbf{y})}{p_{\mathbf{Y}}}d\mathbf{y}
\end{align}

By equalities shown in \eqref{eq:equality1} and \eqref{eq:equality2}, it is straightforward that \eqref{eq:second_derivative} can be turn into the following,
\begin{align}
h_{ij}=\dfrac{\partial^2 h(\mathbf{X}|\mathbf{Y})}{\partial^2\eta_i\eta_j} = -&\int\int\dfrac{1}{p_{\mathbf{XY}}}
[f_i(\mathbf{x,y})-\tilde{g}_i(\mathbf{y})p_{\mathbf{X|Y}}]\times\notag\\
&[f_j(\mathbf{x,y})-\tilde{g}_j(\mathbf{y})p_{\mathbf{X|Y}}]d\mathbf{x}d\mathbf{y}
\end{align}

The matrix $H=[h_{ij}],~i,j\in[0,2^k-1]$ characterizes the Hessian matrix the conditional entropy $h(\mathbf{X|Y})$. To prove the concavity, we need to show $H$ is non-positive definite.
Define a non-zero real row vector $\mathbf{c}\in R^{2^k}$, then we need to form $\mathbf{c}H\mathbf{c}^T$ as follows and show that it is non-positive.
\begin{align}
\mathbf{c}H\mathbf{c}^T=-\int\int\dfrac{1}{p_{\mathbf{XY}}}&
\sum_{i=0}^{2^k-1}\sum_{j=0}^{2^k-1} c_ic_j[f_i(\mathbf{x,y})-\tilde{g}_i(\mathbf{y})p_{\mathbf{X|Y}}]\notag\\
&[f_j(\mathbf{x,y})-\tilde{g}_j(\mathbf{y})p_{\mathbf{X|Y}}]d\mathbf{x}d\mathbf{y}\notag\\
=-\int\int\dfrac{1}{p_{\mathbf{XY}}}
[&\sum_{i=0}^{2^k-1}c_i(f_i(\mathbf{x,y})-\tilde{g}_i(\mathbf{y})p_{\mathbf{X|Y}})]^2d\mathbf{x}d\mathbf{y}\notag\\
&\leq 0
\end{align}

Now that we showed the concavity of the conditional entropy with respect to $\eta_i$, we only need to find the optimal solution. The formulation is defined in \eqref{eq:lagrange}, where $\lambda$ is the Lagrange multiplier.
\begin{align} \label{eq:lagrange}
L = h(\mathbf{X}|\mathbf{Y}) - \lambda\sum_{i=0}^{2^k-1} \eta_i
\end{align}

\noindent by taking derivative with respect to $\eta_i$, we may deduce the following,
\begin{align}
\dfrac{\partial L}{\partial \eta_i} = &-\int\int f_i(\mathbf{x,y})\log p_{\mathbf{XY}}d\mathbf{x}d\mathbf{y}\notag\\
&+ \int \tilde{g}_i(\mathbf{y})\log p_{\mathbf{Y}}d\mathbf{y} - \lambda\notag\\
&=-\int\int f_i(\mathbf{x,y})\log p_{\mathbf{X|Y}}d\mathbf{x}d\mathbf{y} - \lambda
\end{align}

\noindent where the last equality is due to $\tilde{g}_i(\mathbf{y})=\int f_i(\mathbf{x,y})d\mathbf{x}$.
One may find the optimal solution by solving $\partial L/\partial\eta_i=0$ for all $i\in[0,2^k-1]$, which results in showing that $-\int\int [f_i(\mathbf{x,y})-f_j(\mathbf{x,y})]\log p_{\mathbf{X|Y}}d\mathbf{x}d\mathbf{y}=0$, for all $i,j\in[0,2^k-1]$.
In order to find the joint Gaussian density $f_i(\mathbf{x,y})$, observe that we should compute the exponent $[\mathbf{xy}]\Sigma^{-1}_{\mathbf{xy}}[\mathbf{xy}]'$. Since, we are dealing with a latent Gaussian tree, the structure of $U=\Sigma^{-1}_{\mathbf{xy}}$ can be summarized into four blocks as follows \cite{chernoff}. $U_\mathbf{x}$ that has diagonal and off-diagonal entries $u_{x_i}$ and $u_{x_ix_j}$, respectively, and not depending on the edge-signs;
$U_{\mathbf{xy}}$, with nonzero elements $u_{x_iy_j}$ showing the edges between $x_i$ and particular $y_j$ and depending on correlation signs;
$[U_{\mathbf{xy}}]^T$;
$U_{\mathbf{y}}$, with nonzero off diagonal elements $u_{y_iy_j}$  that are a function of edge-sign values, while the diagonal elements $u_{y_i}$ are independent of edge-sign values.
One may show,
\begin{align}
[\mathbf{xy}]\Sigma^{-1}_{\mathbf{xy}}[\mathbf{xy}]'
&= [\sum_{i=1}^n x_i^2 u_{x_i} + \sum_{i=1}^k y_i^2 u_{y_i}]\notag\\
&+ 2[\sum_{n^x_{y_1}} x_iy_1u_{x_iy_1}+...+\sum_{n^x_{y_k}} x_iy_ku_{x_iy_k}]\notag\\
&+ 2\sum_{(i,j)\in E_X} x_ix_ju_{x_ix_j} + 2\sum_{(i,j)\in E_Y} y_iy_ju_{y_iy_j}\notag\\
&= t + 2\sum_{i=1}^k p_i + 2s + 2\sum_{(i,j)\in E_Y} y_iy_ju_{y_iy_j}
\end{align}

\noindent where $n^x_{y_i}$ are the observed neighbors of $y_i$, and $E_Y$ is the edge set corresponding only to hidden nodes, i.e., those hidden nodes that are adjacent to each other.
$E_X$ can be defined similarly, with $s=\sum_{(i,j)\in E_X} x_ix_ju_{x_ix_j}$. 
Also $p_j=\sum_{n^x_{y_j}} x_iy_ju_{x_iy_j}$.
Suppose $f_i(\mathbf{x,y})$ and $f_j(\mathbf{x,y})$ are different at $l$ sign values $\{i_1,...,i_l\}\in L$. Let's write,
\begin{align}
\sum_{(i,j)\in E_Y} y_iy_ju_{y_iy_j}& = \sum_{\substack{{(i,j)\in E_Y}\\{{i,j\in L}}~or~{i,j\notin L}}} y_iy_ju_{y_iy_j}\notag\\
&+ \sum_{\substack{{(i,j)\in E_Y}\\{i~or~j\in L}}} y_iy_ju_{y_iy_j}\notag\\
& = q + q'
\end{align}

Hence, we divide the summation $\sum_{(i,j)\in E_Y} y_iy_ju_{y_iy_j}$ into two parts $q$ and $q'$.
Suppose $\eta_i=1/2^k$ for all $i\in[0,2^k-1]$.
We may form $f_i(\mathbf{x,y})-f_j(\mathbf{x,y})$ as follows,
\begin{align*}
f_i(\mathbf{x,y})-f_j(\mathbf{x,y})\propto &e^{-t/2+s+q+\sum_{i\notin L} p_i}\notag\\
&\times[e^{q'+\sum_{i\in L} p_i}-e^{-q'-\sum_{i\in L} p_i}]
\end{align*}

By negating all $y_{i_1},...,y_{i_l}$ into $-y_{i_1},...,-y_{i_l}$, it is apparent that $t$, $\sum_{i\notin L} p_i$, and $s$ do not change. Also, the terms in $q$ either remain intact or doubly negated, hence, overall $q$ remains intact also. 
However, by definition, $p_i,i\in L$ will be negated, hence overall the sum $\sum_{i\in L} p_i$ will be negated. The same thing holds true for $q'$, since exactly one variable $y_i$ or $y_j$ in the summation, will change its sign, so $q'$ also will be negated.
Overall, we can see that by negating $y_{i_1},...,y_{i_l}$, we will negate $f_i-f_j$.
It remains to show that such negation does not impact $p_{\mathbf{X|Y}}$. Note that since $p_{\mathbf{XY}}$ includes all $2^k$ sign combinations and all of $f_i(\mathbf{x,y})$ are equi-probable since we assumed $\eta_i=1/2^k$ so $p_{\mathbf{XY}}$ is symmetric with respect to $\eta_i$, and such transformation on $y_{i_1},...,y_{i_l}$ does not impact the value of $p_{\mathbf{XY}}$, since by such negation we simply switch the position of certain Gaussian terms $f_i(\mathbf{x,y})$ with each other.

For $p_{\mathbf{y}}$, we should first compute the term $\mathbf{y}\Sigma_{\mathbf{y}}^{-1}\mathbf{y}'$. We know $\Sigma_{\mathbf{y}}=B\Sigma'_{\mathbf{y}} B$, so $\Sigma^{-1}_Y=B^{-1}\Sigma'^{-1}_{\mathbf{y}} B^{-1}=B\Sigma'^{-1}_{\mathbf{y}} B$ (note, $\Sigma_{\mathbf{y}}$ does not necessarily induce a tree structure). 
We have,
\begin{align*}
\mathbf{y}\Sigma_{\mathbf{y}}^{-1}\mathbf{y}'= \sum_{i=1}^k w_{ii} y_i^2 + 2\sum_{i,j\& i<j} w_{ij} y_iy_jb_ib_j 
\end{align*}

From this equation, we may interpret the negation of $y_{i_1},...,y_{i_l}$, simply as negation of $b_{i_1},...,b_{i_l}$. Hence, since $p_{\mathbf{y}}$ includes all sign combinations, hence, such transformation only permute the terms $\tilde{g}_i(\mathbf{y})$, so $p_{\mathbf{y}}$ remains fixed.
Hence, overall $p_{\mathbf{X|Y}}$ remains unaltered.
As a result, we show that for any given point in the integral $\int\int (f_i(\mathbf{x,y})-f_j(\mathbf{x,y}))\log p_{\mathbf{X|Y}}d\mathbf{x}d\mathbf{y}$ we can find its negation, hence making the integrand an odd function, and the corresponding integral zero.
Hence, making the solution $\eta_i=1/2^k$, for all $i\in[0,2^k-1]$ an optimal solution.

The only thing remaining is to show that from $\eta_i=1/2^k$ we may conclude that $\pi_j=1/2$ for all $j\in[1,k]$.
By definition, we may write,
\begin{align*}
\eta_i=\prod_{j=1}^k \pi_j^{b_{ji}}(1-\pi_j)^{1-b_{ji}}
\end{align*}

\noindent where $b_{ji}\in B_i$.
Assume all $\eta_i=1/2^k$. Consider $\eta_1$ and find $\eta_{i^*}$ such that the two are different in only one expression, say at the $l$-th place.
Since, all $\eta_i$ are equal, one may deduce $1-\pi_l=\pi_l$ so $\pi_l=1/2$. Note that such $\eta_{i^*}$ can always be found since $\eta_i$'s are covering all possible combinations of $k$-bit vector. 
Now, find another $\eta_{j^*}$, which is different from $\eta_1$ at some other spot, say $l'$, again using similar arguments, we may show $\pi_{l'}=1/2$.
This can be done $k$ times to show that, if all $\eta_i=1/2^k$, then $\pi_1=...=\pi_k=1/2$.
This completes the proof.

\section{Proof of Theorem \ref{thm:achievability_basic}} \label{app:achievability_basic}
The signal model can be directly written as follows,
\begin{align} \label{new_signal}
\mathbf{Y}^{(l)} = A_{\mathbf{B}^{(l,l+1)}} \mathbf{Y}^{(l+1)} + \mathbf{Z}^{(l)}
\end{align}

\begin{figure} [h!]
\centering
\includegraphics[width=0.75\columnwidth]{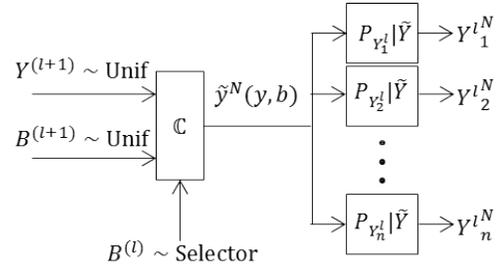} 
\caption{Construction of the joint density $\gamma_{(\mathbf{Y}^{(l)})^N,\mathbf{Y}^{(l+1)},\mathbf{B}^{(l+1)}}$}\label{fig:encoding_basic} 
\end{figure}

Here, we show the codebook generation scheme to generate $\mathbf{Y}^{(l)}$ from $\mathbf{Y}^{(l+1)}$.
Note that $\mathbf{Y}^{(l)}$ is a vector consisting of the variables $Y_i^{(l)}$. Also, $\mathbf{Y}^{(l+1)}$ is a vector consisting of variables $Y_i^{(l+1)}$.
The proof relies on the procedure taken in \cite{cuff}. Note that our scheme should satisfy the following constraints,

\begin{tabular}{l l}
$1) ({Y}^{(l)}_i)^N\perp ({Y}^{(l)}_j)^N|\tilde{\mathbf{Y}}^{(l+1)}~~ (i\neq j)$\\
$2) (\mathbf{Y}^{(l)})^N\perp \mathbf{B}^{(l+1)}$\\
$3) P_{(\mathbf{Y}^{(l)})^N} = \prod_{t=1}^N P_{\mathbf{Y}^{(l)}}(\mathbf{y}^{(l)}_t)$\\ 
$4) |\mathbf{Y}^{(l+1)}|= 2^{NR_{\mathbf{Y}^{(l+1)}}}$\\
$5) |\mathbf{B}^{(l+1)}|= 2^{NR_{\mathbf{B}^{(l+1)}}}$\\
$6) ||q_{(\mathbf{Y}^{(l)})^N}-\prod_{t=1}^N P_{\mathbf{Y}^{(l)}}(\mathbf{y}^{(l)}_t)||_{TV}<\epsilon$
\end{tabular}

\noindent where the first constraint is due to the conditional independence assumption characterized in the signal model \eqref{new_signal}. The second one is to capture the intrinsic ambiguity of the latent Gaussian tree to capture the sign information. Condition $3)$ is due to independence of joint densities $P_{\mathbf{Y}^l}(\mathbf{Y}^l_t)$ at each time slot $t$. Conditions $4)$ and $5)$ are due to corresponding rates for each of the inputs $\mathbf{Y}^{(l+1)}$ and $\mathbf{B}^{(l+1)}$. And finally, condition $6)$ is the synthesis requirement to be satisfied.
First, we generate a codebook $\mathcal{C}$ of $\tilde{y}^N$ sequences, with indices $y\in C_Y=\{1,2,...,2^{NR_{\mathbf{Y}^{(l+1)}}}\}$ and $b\in C_B=\{1,2,...,2^{NR_{\mathbf{B}^{(l+1)}}}\}$ according to the explained procedure in Algorithm \ref{alg:codebook}.
The codebook $\mathcal{C}$ consists of all combinations of the sign and latent variables codewords, i.e., $|\mathcal{C}| = |C_Y|\times |C_B|$.
We construct the joint density $\gamma_{(\mathbf{Y}^{(l)})^N,\mathbf{Y}^{(l+1)},\mathbf{B}^{(l+1)}}$ as depicted by Figure \ref{fig:encoding_basic},

The indices $y$ and $b$ are chosen independently and uniformly from the codebook $\mathcal{C}$.
As can be seen from Figure \ref{fig:encoding_basic}, for each $\mathbf{B}^{(l)}_t = \mathbf{b}^{(l)}_t$ the channel $P_{{Y}^l|\tilde{Y}}$ is in fact consists of $n$ independent channels $P_{{Y}^l_i|\tilde{Y}},~i\in\{1,2,...,n\}$. The joint density is as follows,
\begin{align*}
\gamma_{(\mathbf{Y}^{(l)})^N,\mathbf{Y}^{(l+1)},\mathbf{B}^{(l+1)}}=\dfrac{1}{|C_Y||C_B|}[\prod_{t=1}^N P_{\mathbf{Y}^l}(\mathbf{Y}^l_t|\tilde{y}_t(y,b))]
\end{align*}

Note that $\gamma_{(\mathbf{Y}^{(l)})^N,\mathbf{Y}^{(l+1)},\mathbf{B}^{(l+1)}}$ already satisfies the constraints $1)$, $4)$, and $5)$ by construction.
Next, we need to show that it satisfies the constraint $6)$. The marginal density $\gamma_{(\mathbf{Y}^{(l)})^N}$ can be deduced by the following,
\begin{align*}
\gamma_{(\mathbf{Y}^{(l)})^N}=\dfrac{1}{|C_Y||C_B|}\sum_{y\in C_Y}\sum_{b\in C_B}[\prod_{t=1}^N P_{\mathbf{Y}^{(l)}}(\mathbf{Y}^{(l)}_t|\tilde{y}_t(y,b))]
\end{align*}
We know if $R_{\mathbf{B}^{(l+1)}}+R_{\mathbf{Y}^{(l+1)}} \geq I[\mathbf{Y}^{(l+1)},\mathbf{B}^{(l+1)};\mathbf{Y}^{(l)}|\mathbf{B}^{(l)}]$, then by soft covering lemma \cite{cuff} we have,
\begin{align} \label{eq:6}
\lim_{n\rightarrow\infty} E||\gamma_{(\mathbf{Y}^{(l)})^N}-\prod P_{\mathbf{Y}^{(l)}}||_{TV} = 0
\end{align}

\noindent which shows that $\gamma_{(\mathbf{Y}^{(l)})^N}$ satisfies constraint $6)$. For simplicity of notations we use $\prod P_{\mathbf{Y}^{(l)}}$ instead of $\prod_{t=1}^N P_{\mathbf{Y}^{(l)}}(\mathbf{Y}^{(l)}_t)$, since it can be understood from the context.
Next, let's show that $\gamma_{(\mathbf{Y}^{(l)})^N}$, \textit{nearly} satisfies constraints $2)$ and satisfies $3)$. We need to show that as $N$ grows the synthesized density $\gamma_{(\mathbf{Y}^{(l)})^N,\mathbf{B}^{(l+1)}}$ approaches $\dfrac{1}{|C_B|}\prod P_{\mathbf{Y}^{(l)}}$, in which the latter satisfies both $2)$ and $3)$. In particular, we need to show that the total variation 
$E||\gamma_{(\mathbf{Y}^{(l)})^N,\mathbf{B}^{l+1}}-\dfrac{1}{|C_B|}\prod P_{\mathbf{Y}^{(l)}}||$ vanishes as $N$ grows.
After taking several algebraic steps similar to the ones in \cite{cuff}, we should equivalently show that the following term vanishes, as $N\rightarrow\infty$,
\begin{align} \label{eq:nearly}
\dfrac{1}{|C_B|}\sum_{b\in C_B}E||\gamma_{(\mathbf{Y}^{(l)})^N,\mathbf{B}^{l+1}=\mathbf{b}}-\prod P_{\mathbf{Y}^{(l)}}||_{TV}
\end{align}

Note that given any fixed $b\in C_B$ the number of Gaussian codewords is $|C_Y|=2^{NR_{\mathbf{Y}^{(l+1)}}}$. 
Also, one can check by the signal model defined in \eqref{new_signal} that the statistical properties of the output vector $\mathbf{Y}^{(l)}$ given any fixed sign value $b\in C_B$ does not change. 
Hence, for sufficiently large rates, i.e., $R_{\mathbf{Y}^{(l+1)}}\geq I[\mathbf{Y}^{(l+1)};\mathbf{Y}^{(l)}|\mathbf{B}^{(l)}]$, and by soft covering lemma, the term in the summation in \eqref{eq:nearly} vanishes as $N$ grows.
So overall the term shown in \eqref{eq:nearly} vanishes.
This shows that in fact $\gamma_{(\mathbf{Y}^{(l)})^N}$ \textit{nearly} satisfies the constraints $2)$ and $3)$.
Hence, let's construct another distribution using $\gamma_{(\mathbf{Y}^{(l)})^N,\mathbf{Y}^{(l+1)},\mathbf{B}^{(l+1)}}$. Define,
\begin{align*}
q_{(\mathbf{Y}^{(l)})^N,\mathbf{Y}^{(l+1)},\mathbf{B}^{(l+1)}}=\dfrac{1}{|C_B|}(\prod P_{\mathbf{Y}^{(l)}})\gamma_{\mathbf{Y}^{(l+1)}|(\mathbf{Y}^{(l)})^N,\mathbf{B}^{(l+1)}}
\end{align*}

It is not hard to see that such density satisfies $1)-5)$. 
We only need to show that it satisfies $6)$ as well. We have,
\begin{align}
&||q_{(\mathbf{Y}^{(l)})^N}-\prod P_{\mathbf{Y}^{(l)}}||_{TV}\notag\\
&\leq ||q_{(\mathbf{Y}^{(l)})^N}-\gamma_{(\mathbf{Y}^{(l)})^N}||_{TV} + ||\gamma_{(\mathbf{Y}^{(l)})^N}-\prod P_{\mathbf{Y}^{(l)}}||_{TV}\notag\\
&\leq ||q_{(\mathbf{Y}^{(l)})^N,\mathbf{Y}^{(l+1)},\mathbf{B}^{(l+1)}}-\gamma_{(\mathbf{Y}^{(l)})^N,\mathbf{Y}^{(l+1)},\mathbf{B}^{(l+1)}}||_{TV} + \epsilon_N\label{eq:2nd}\\
&=||q_{(\mathbf{Y}^{(l)})^N,\mathbf{B}^{(l+1)}}-\gamma_{(\mathbf{Y}^{(l)})^N,\mathbf{B}^{(l+1)}}||_{TV} + \epsilon_N\label{eq:4th}\\
&=||\dfrac{1}{|C_B|}(\prod P_{\mathbf{Y}^{(l)}})-\gamma_{(\mathbf{Y}^{(l)})^N,\mathbf{B}^{(l+1)}}||_{TV} +  \epsilon_N \label{eq:condition6}
\end{align}

\noindent where $\epsilon_N=||\gamma_{(\mathbf{Y}^{(l)})^N}-\prod P_{\mathbf{Y}^{(l)}}||_{TV}$.
Both terms in \eqref{eq:condition6} vanish as $N$ grows, due to \eqref{eq:nearly} and \eqref{eq:6}, respectively. Note that, \eqref{eq:2nd} is due to \cite[Lemma V.I]{cuff}. Also, \eqref{eq:4th} is due to \cite[Lemma V.II]{cuff}, by considering the terms $q_{(\mathbf{Y}^{(l)})^N,\mathbf{Y}^{(l+1)},\mathbf{B}^{(l+1)}}$ and $\gamma_{(\mathbf{Y}^{(l)})^N,\mathbf{Y}^{(l+1)},\mathbf{B}^{(l+1)}}$ as the outputs of a unique channel specified by $\gamma_{\mathbf{Y}^{(l+1)}|(\mathbf{Y}^{(l)})^N,\mathbf{B}^{(l+1)}}$, with inputs $p_{(\mathbf{Y}^{(l)})^N,\mathbf{B}^{(l+1)}}$ and $\gamma_{(\mathbf{Y}^{(l)})^N,\mathbf{B}^{(l+1)}}$, respectively.

Finally, note that we synthesize each $\mathbf{Y}^{(l)}$ for a given $\mathbf{B}^{(l)}=\mathbf{b}$. Hence, to obtain the overall statistics we have $q_{\mathbf{Y}^{(l)}}=\sum_{b}q_{\mathbf{Y}^{(l)}|\mathbf{B}^{(l)}}p(\mathbf{B}^{(l)}=\mathbf{b})$, where the summation is over all possible sign combinations for layer $\mathbf{Y}^{(l)}$, which equals to $2^{k_l}$. Certainly, this number becomes exponentially large if $k_l$ is large. However, note that as $N\rightarrow\infty$ each synthesized output (for each given $\mathbf{B}^{(l)}=\mathbf{b}$) become arbitrarily close to zero. Hence, overall $q_{\mathbf{Y}^{(l)}}$ becomes arbitrarily close to the desired statistics. This is also the case for the overall latent Gaussian tree, i.e., for $L$ capturing the total number of layers, at each layer we can generate an output with vanishing total variation distance from the desired statistics, hence overall the final output statistics becomes arbitrarily close to the desired output statistics.

This completes the achievability proof.

\section{Proof of Lemma \ref{lem:achievability_general2}} \label{app:achievability_general2}

First, we need to change the latent tree structure in a way similar to Figure \ref{fig:internal3}.
We start from the standard latent structure, and at each layer we seek for those latent nodes that are at the same layer and they are neighbors.
For each pair of adjacent nodes, we move the one that is further away from the top layer to a new added layer below the current one.
Hence, make a new layer of latent nodes.
We iterate this step until we reach the bottom layer.
This way, we face different groups of observables being synthesized at different layers.

Define $\mathbf{X}^{(l)}$, $\mathbf{Y}^{(l)}$ and $\mathbf{B}^{(l)}$ as the set of observables, latent nodes and sign variables at layer $l$, respectively.
In this new setting layer $l=0$ defines the observable layer, which only consists of remaining output variables, with no latent nodes.
If the rates at each layer satisfy the inequalities in  \eqref{eq:thm_achievability_basic}, then by Theorem \ref{thm:achievability_basic} we know that as $N$ increases, the simulated density $q_{(\mathbf{X}^{(l)})^N,(\mathbf{Y}^{(l)})^N}$ approaches to the desired density $\prod p_{(\mathbf{X}^{(l)}),(\mathbf{Y}^{(l)})}$.
Suppose the first set of outputs are generated at layer $L'$, then we know $\mathbf{X}=\bigcup_{l=0}^{L'} \mathbf{X}^{(l)}$.
Each observable node ${X}_i^{(l)}$, for $l<L'$ has a latent ancestor at each layer $l<l'\leq L'$. 
We define $\mathbf{Y}'$ as the union of latent nodes containing all those latent ancestors.
Basically, the vector $\mathbf{Y}'$ includes all the latent nodes ${Y}_j^{(l)}$ for $1\leq l\leq L'$.
We define $\mathbf{B}'$, similarly, i.e., those sign inputs related to the nodes in the set $\mathbf{Y}'$.
With slightly abuse of notation, define $\tilde{\mathbf{Y}}=\{\mathbf{Y}',\mathbf{B}'\}$, and $\tilde{\mathbf{Y}}^{(l)}=\{\mathbf{Y}^{(l)},\mathbf{B}^{(l)}\}$, for all possible layers $l$.
The scheme looks exactly as discussed previously, except that this time we need to keep track of corresponding generated outputs at each layer and match them together.
In particular, consider the generated outputs $(\mathbf{X}^{(0)})^N$, which lie at the bottom layer. Each output is generated using a particular input vector $(\mathbf{Y}^{(1)})^N$, which in turn along with other possible outputs $(\mathbf{X}^{(1)})^N$ are generated by a unique input codeword $(\mathbf{Y}^{(2)})^N$ that lie at the second layer.
This procedure moves from the bottom to the top layer, in order to match each generated output at the bottom layer with the correct output vectors at other layers.
Note that the sign information will be automatically taken care of, since similar to the previous cases, at each layer $l+1$ and given each realization of the sign vector $\mathbf{B}^{(l)}=\mathbf{b}^{(l)}$, the input vector $\mathbf{Y}^{(l+1)}$ will become Gaussian.
We only need to show that the synthesize density regarding to such formed joint vectors approaches to the desired output density, as $N$ grows.

By the underlying structure of latent tree, one may factorize the joint density $q_{\mathbf{X}^N,{\tilde{\mathbf{Y}}}^N}=q_{(\mathbf{X}^{(L')})^N,(\tilde{\mathbf{Y}}^{(L')})^N}\prod_{l=0}^{L'-1}q_{(\mathbf{X}^{(l)})^N|(\tilde{\mathbf{Y}}^{(l+1)})^N}$.
Note that the desired joint density $p_{\mathbf{X},\tilde{\mathbf{Y}}}$ also induces the same latent Gaussian tree, hence, we may write, $p_{\mathbf{X}^N,{\tilde{\mathbf{Y}}}^N}=p_{(\mathbf{X}^{(L')})^N,(\tilde{\mathbf{Y}}^{(L')})^N}\prod_{l=0}^{L'-1}p_{(\mathbf{X}^{(l)})^N|(\tilde{\mathbf{Y}}^{(l+1)})^N}$.
However, by our synthesis scheme shown in Figure \ref{fig:encoding_basic}, one may argue that $\prod_{l=0}^{L'-1}q_{(\mathbf{X}^{(l)})^N|(\tilde{\mathbf{Y}}^{(l+1)})^N}=\prod_{l=0}^{L'-1}p_{(\mathbf{X}^{(l)})^N|(\tilde{\mathbf{Y}}^{(l+1)})^N}=\prod_{l=0}^{L'-1}\prod p_{\mathbf{X}^{(l)}_t|\tilde{\mathbf{Y}}^{(l+1)}_t}$.
By summing out $(\mathbf{B}^{(L')})^N$ from both densities $p_{\mathbf{X}^N,{\tilde{\mathbf{Y}}}^N}$ and $q_{\mathbf{X}^N,{\tilde{\mathbf{Y}}}^N}$, we may replace $p_{(\mathbf{X}^{(L')})^N,(\tilde{\mathbf{Y}}^{(L')})^N}$ with $p_{(\mathbf{X}^{(L')})^N,({\mathbf{Y}}^{(L')})^N}$ and $q_{(\mathbf{X}^{(L')})^N,(\tilde{\mathbf{Y}}^{(L')})^N}$ with $q_{(\mathbf{X}^{(L')})^N,({\mathbf{Y}}^{(L')})^N}$, since only these terms in the equations depend on the sign vector at layer $L'$, i.e., $(\mathbf{B}^{(L')})^N$.
Now, by previous arguments for the synthesized and desired density at layer $L'$, we know that the total variation distance $||q_{(\mathbf{X}^{(L')})^N,(\mathbf{Y}^{(L')})^N}-\prod p_{\mathbf{X}^{(L')}_t,\mathbf{Y}^{(L')}_t}||_{TV}$ goes to zero as $N$ grows.
Hence, one may simply deduce that $||q_{\mathbf{X}^N,{\tilde{\mathbf{Y}}}^N/(\mathbf{B}^{(L')})^N}-\prod p_{\mathbf{X}_t,{\tilde{\mathbf{Y}}}_t/\mathbf{B}^{(L')}_t}||_{TV}=||(q_{(\mathbf{X}^{(L')})^N,(\mathbf{Y}^{(L')})^N}-\prod p_{\mathbf{X}^{(L')}_t,\mathbf{Y}^{(L')}_t})\prod_{l=0}^{L'-1}\prod p_{\mathbf{X}^{(l)}_t|\tilde{\mathbf{Y}}^{(l+1)}_t}||_{TV}$ goes to zero as $N$ grows.
Due to \cite[Lemma V.I]{cuff}, we know $||q_{\mathbf{X}^N}-\prod p_{\mathbf{X}_t}||_{TV}\leq ||q_{\mathbf{X}^N,{\tilde{\mathbf{Y}}}^N/(\mathbf{B}^{(L')})^N}-\prod p_{\mathbf{X}_t,{\tilde{\mathbf{Y}}}_t/\mathbf{B}^{(L')}_t}||_{TV}<\epsilon$, and as $N$ grows.
This completes the proof.

\end{document}